\documentclass[useAMS]{mn2e}
\usepackage{graphicx}

\title[Dust formation by the colliding-wind binary WR140]
{Orbitally modulated dust formation by the WC7+O5 colliding-wind binary WR\,140}
\author[P. M. Williams et al.]
       {P. M. Williams$^1$\thanks{Email: pmw@roe.ac.uk}, 
        S. V. Marchenko$^2$, A. P. Marston$^{3,4}$, A. F. J. Moffat$^5$, \and
        W. P. Varricatt$^{6}$, S. M. Dougherty$^7$, M. R. Kidger$^{4,8}$, L. Morbidelli$^9$, 
        M. Tapia$^{10}$\\
   $^1$Institute for Astronomy, Scottish Universities Physics Alliance, 
     University of Edinburgh, Royal Observatory, Edinburgh EH9 3HJ\\
   $^2$Dept of Physics and Astronomy, Western Kentucky University, 
     1906 College Heights Blvd \#11077, Bowling Green, KY 42101, USA\\
   $^3$SIRTF Science Center, IPAC, Caltech, Mail Stop 314-6, Pasadena, CA 91125, USA\\
   $^4$Herschel Science Centre, European Space Astronomy Centre, Villafranca del Castillo,
     P.O.Box - Apdo.50727, 28080 Madrid, Spain \\
   $^5$D\'epartement de physique, Universit\'e de Montr\'eal, C.P. 6128, Succ. Centre-Ville, 
     Montr\'eal, QC, H3C 3J7, Canada \\
   $^6$Joint Astronomy Centre, 660 N. A`oh\=ok\=u Place, Hilo,  HI 96720, USA\\   
   $^7$National Research Council of Canada, Herzberg Institute for Astrophysics, Dominion 
      Radio Astrophysical Observatory, P O Box 248, \\ Penticton, B.C. V2A 6J9, Canada\\
   $^8$Ingenieria y Servicios Aeroespaciales SA, ESAC, Villafranca del Castillo, 
      P.O.Box - Apdo.50727, 28080 Madrid, Spain\\
   $^9$INAF - Osservatorio Astrofisico di Arcetri, Largo E. Fermi 5, I-50125 Firenze, Italy\\
   $^{10}$Universidad Nacional Aut\'{o}noma de M\'{e}xico, Instituto de Astronom\'{i}a,
     Apartado Postal 877, Ensenada B.C., Mexico\\
}

\date{Accepted 2009 February 19.
      Received 2009 February 16;
      in original form 2008 Dec 23}

\pagerange{\pageref{firstpage}--\pageref{lastpage}}
\pubyear{2009}

\begin{document}

\maketitle

\label{firstpage}

\begin{abstract}
We present high-resolution infrared (2--18 $\umu$m) images of the archetypal 
periodic dust-making Wolf-Rayet binary system WR\,140 (HD 193793) taken between 
2001 and 2005, and multi-colour ($J$ -- [19.5]) photometry observed between 
1989 and 2001. The images resolve the dust cloud formed by WR\,140 in 2001, 
allowing us to track its expansion and cooling, while the photometry allows 
tracking the average temperature and total mass of the dust. The combination 
of the two datasets constrains the optical properties of the dust, and suggest 
that they differ from those of the dust made by the WC9 dust-makers, including 
the classical `pinwheel', WR\,104. Photometry of individual dust emission 
features shows them to be significantly redder in ($nbL^{\prime}$--[3.99]), 
but bluer in ([7.9]--[12.5]), than the binary, as expected from the spectra 
of heated dust and the stellar wind of a Wolf-Rayet star. 
The most persistent dust features, two concentrations at the ends of a `bar' 
of emission to the south of the star, were observed to move with constant 
proper motions of $324\pm8$ and $243\pm7$ mas y$^{-1}$. 
Longer wavelength (4.68-$\umu$m and 12.5-$\umu$m) images shows dust emission 
from the corresponding features from the previous (1993) periastron passage 
and dust-formation episode, showing that the dust expanded freely in a 
low-density void for over a decade, with dust features repeating from one 
cycle to the next. A third persistent dust concentration to the east of the 
binary (the `arm') was found to have a proper motion $\sim$ 320 mas y$^{-1}$, 
and a dust mass about one-quarter that of the `bar'. 
Extrapolation of the motions of the concentrations back to the binary suggests 
that the eastern `arm' began expansion 4--5 months earlier than those in the 
southern `bar', consistent with the projected rotation of the binary axis and 
wind-collision region (WCR) on the sky. Comparison of model dust images and 
the observations constrain the intervals when the WCR was producing sufficiently 
compressed wind for dust nucleation in the WCR, and suggests that the 
distribution of this material was not uniform about the axis of the WCR, but 
more abundant in the following edge in the orbital plane.

\end{abstract}

\begin{keywords}
stars: circumstellar matter --- infrared: stars --- stars: individual: 
WR\,140 --- stars: Wolf-Rayet. 
\end{keywords}

\newpage

\section{Introduction}

\noindent In 1977, the infrared (IR) flux from HD\,193793 (WR\,140) was 
observed to have risen by almost an order of magnitude 
(e.g. $\Delta L^{\prime} \simeq 2.4$) in less than a year and to  
have developed a spectral energy distribution (SED) indicating formation of 
circumstellar dust (Williams et al. 1978; Hackwell, Gehrz \& Grasdalen 1979). 
Although infrared SEDs characteristic of heated circumstellar dust had 
previously been observed from WC8--9 type Wolf-Rayet stars (e.g., Allen, 
Harvey \& Swings 1972, Gehrz \& Hackwell 1974), HD\,193793 was the first 
WR star to show evidence for the formation of circumstellar dust on a 
short time-scale. Subsequent observations showed the emission to fade as 
the dust cooled and then, in 1985, to rise again in another dust-formation 
episode (Williams et al. 1990, hereafter Paper~1). From the IR light curves, 
a photometric period of $2900\pm10$d was derived, leading to a solution 
of the radial velocities which showed WR\,140 to be a high-eccentricity 
($e=0.84$) WC7+O5\footnote{Spectral types between O4--5 and O6 have been 
assigned to the primary in different studies. We re-examined the 
classification by measuring equivalent widths of the 
He\,{\sc i} $\lambda$4471 and He\,{\sc ii} $\lambda$4542 lines, 
and derived the classification ratio (Conti \& Alschuler 1971) 
log $W^{\prime} = \log W(4471) - \log W(4542) = -0.41\pm0.05$, 
which gives O5.5. This is later than the O4--5 derived from the 
optical spectrum by Lamontagne, Moffat \& Seggewiss (1984) and from 
the UV spectrum in Paper~1, but is consistent with the near equality 
of the He\,{\sc i} (+ He\,{\sc ii}) $\lambda$4025 and He\,{\sc ii} 
$\lambda$4200 lines (cf. Walborn \& Fitzpatrick 1990). We adopt O5.} 
spectroscopic binary. The episodes of dust formation appear to coincide  
with periastron passage (Paper~1). 

In the interim, rises in IR emission indicative of episodes of dust 
formation were observed from two other WR stars, WR\,48a (WC8--9) in 1978-79 
(Danks et al. 1983) and WR\,137 (WC7 + O) in 1983-4 (Williams et al. 1985),  
showing that WR\,140 was not unique, but rather a prototype of a class of 
episodic dust-making WR stars, of which we now know seven (Williams 2002). 

The particular interest and problem posed by the episodic dust-making WR 
stars is the great difficulty of making dust in their winds, as pointed 
out in the case of WR\,140 by Hackwell et al. (1979): the stellar winds are 
not sufficiently dense to allow homonuclear grain growth. The same problem 
was discussed in respect of the `classical' dust-makers amongst the WC8--9 
stars (Williams, van der Hucht, \& Th\'e 1987). Zubko (1998) found that 
carbon grains having a drift velocity relative to the WC8--9 wind could 
grow via collision with carbon ions, but the question of grain nucleation 
remains open. 
A chemical kinetic study of the formation of molecular species and carbon 
grain precursors in WC9 winds by Cherchneff et al. (2000) found that 
only C$_2$ was formed in useful amounts, and even that required 
significantly (factor $10^3$) higher densities than expected in the 
regions of isotropic, smooth WR winds where dust grains are observed. 
Evidently, the observation of dust formation by WR stars requires that 
their winds contain regions of significantly higher density than expected 
in an isotropic smooth wind. 

Large-scale, high-density structures are provided by compression of stellar 
winds in strong shocks formed where the wind of a WR star collides with 
that of a sufficiently luminous companion in a binary system -- a colliding 
wind binary (CWB) -- as occurs in WR\,140. Usov (1991) suggested that very 
high compression factors ($10^{3} - 10^{4}$) could be produced in WR\,140 
if the heated and compressed wind was able to cool efficiently. 
The link between the dust-formation episodes and binary orbit 
is provided by periodic increases by factor of $\sim$ 40 of the 
{\em pre-shock} wind density at the wind-collision region (WCR) for a brief 
time around periastron passage (Williams 1999). 
Although hydrodynamical modelling by Stevens, Blondin, \& Pollock (1992) 
found the collision region in WR\,140 to be adiabatic, which would give much 
less compression, this does not hold very close to periastron, when the 
compressed wind material is observed to cool through additional emission 
components on low-excitation C\,{\sc iii} and He\,{\sc i} lines
(Marchenko et al. 2003, hereafter MM03; Varricatt, Williams \& Ashok 2004,
hereafter VWA). The cooling in the emission sub-peak on the He\,{\sc i} 
$\lambda$1.083-$\umu$m line exceeds that by the X-rays (VWA). 

Other high-density structures which could aid dust formation are the clumps 
in WR (e.g. Moffat \& Robert 1994) and O (e.g. Eversberg, Lepine \& Moffat 1998) 
star winds. A test of the relative importance of these and the large-scale 
structures can come from imaging the newly-formed dust in the IR. 
Marchenko, Moffat, \& Grosdidier (1998) observed large clumps of dust near 
WR\,137 during its 1997 maximum and Tuthill, Monnier \& Danchi (1999), Monnier, 
Tuthill, \& Danchi (1999) and Tuthill et al. (2008) observed rotating `pinwheel' 
dust structures around WR\,104 (Ve2--45) and WR\,98a (IRAS 17380--3031), 
characteristic of dust formed and emitted in a stream to one side of a binary 
system observed at a relatively low orbital inclination. The observation of 
non-thermal radio emission from the `pinwheel' systems (Monnier et al. 2002a)
provides further support for their binarity; but their stellar components have 
not been resolved, so it has not been possible to relate the positions of the 
stars and WCRs to that of the dust. 

Such a comparison is now possible in the case of WR\,140, whose dust emission 
was mapped by Monnier, Tuthill \& Danchi (2002b, herafter MTD), and whose 
orbit is now fully determined in three dimensions.
The RV orbital study by MM03 found a period, $P=2899\pm1.3$d., in excellent 
agreement with the IR photometric period, and an even greater eccentricity 
($e=0.881\pm0.005$) than those derived previously. 
From high-resolution imaging of the radio emission between phases 0.74 and 
0.97, Dougherty et al. (2005) demonstrated that the WR\,140 system rotates 
clockwise on the sky, and derived an orbital inclination, $i=122\degr\pm5\degr$, 
consistent with and more tightly constrained than that ($i=50\degr\pm15\degr$, 
equivalent to $i=130\degr\pm15\degr$) 
derived by MM03 from modelling the moving sub-peaks on the He\,{\sc i} and 
C\,{\sc iii} line profiles. The binary itself was resolved by Monnier et al. 
(2004), who measured the stellar separation and position angle at phase 0.297. 
Using this and the inclination, Dougherty et al. determined the longitude of 
ascending node ($\Omega=353\degr\pm3\degr$), which, together with the argument 
of periastron ($\omega=46\fdg7\pm1\fdg6$, MM03), completes the definition of 
the orbit. 

The high-resolution near-IR images of WR\,140 observed by MTD in 2001 June 12 
and July 30 showed the dust to lie in several features of different size 
around the central star, and to be expanding with a proper motion of about 
1.1~mas~d$^{-1}$, but with no obvious relation to the orbit.

We therefore extended the imaging obervations of WR\,140 to study the 
evolution of the dust cloud and its relation to the colliding-wind structures 
inferred from the orbit and other observations. 
We also continued the infrared photometry reported in Paper I to confirm the 
periodicity determined from the 1970--1985 data and to improve the definition 
of the light curves around $\phi = 0.14$, which suggest enhanced dust formation, 
perhaps a secondary nucleation episode.

The two datasets are complementary: while the images show positions of the 
dust features and their angular expansion, the photometry allows determination 
of the evolution of the `average' temperature of the dust on account of the wide 
wavelength range and greater frequency of the observations. There is a caveat:
the imaging data follow the 2001 periastron passage, whereas the photometric data 
come mostly from earlier periastron passages. The relation of the two datasets 
relies on the photometeric behaviour repeating periodically, which it appears to 
do (Section \ref{SStart}).

In Section \ref{SObs} we report the infrared photometry in 1989--2001 and 
imaging observations in 2001--5. In Section~\ref{SEvol}, we use the light 
curves and images to describe the evolution of the dust cloud and use the 
positions and fluxes of the dust features to quantify the evolution.  
In Section~\ref{SRelCWB}, we attempt to model the dust images assuming the dust 
forms in the WCR rotating on the sky with the orbital motion and consider the 
influence of model parameters and some of the underlying assumptions on the fit. 
In Section~\ref{SDiscuss}, we discuss the results and the problems arising from 
the timing of the start and finish of the production of dust-forming plasma in 
the WCR and nucleation of the dust.

\begin{table}
\caption{New near-infrared photometry of WR\,140. The phases here and 
in Tables \ref{TmIRP} and \ref{Timages} were calculated using P = 2900~d. 
and $T_{\rmn 0}$ = JD 2446147 (1985.22). Observations made on two 
successive nights are replaced by their average and marked `*'.}
\begin{tabular}{lccccccl}
\hline
Date    &$\phi$& $J$  &  $H$ &  $K$ &$L^{\prime}$ & $nbM$ & Telescope \\
\hline
1989.41 & 0.53 & 5.60 & 5.39 & 5.09 & 4.64 &      & SPM \\ 
1989.65 & 0.56 & 5.60 & 5.42 & 5.10 & 4.70 & 4.32 & UKIRT \\ 
1990.70 & 0.69 & 5.63 & 5.44 & 5.13 &      & 4.40 & UKIRT \\ 
1991.41 & 0.78 & 5.57 & 5.36 & 5.04 & 4.87 &      & SPM \\ 
1991.52 & 0.79 & 5.60 & 5.39 & 5.09 & 4.73 & 4.53 & UKIRT \\ 
1991.79 & 0.83 & 5.59 & 5.39 & 5.08 &      &      & TCS \\ 
1991.84 & 0.83 & 5.59 & 5.37 & 5.11 & 4.78 & 4.49 & UKIRT \\ 
1992.29 & 0.89 & 5.59 & 5.39 & 5.04 & 4.77 &      & TCS \\ 
1992.32 & 0.90 & 5.61 & 5.41 & 5.11 & 4.80 & 4.53 & UKIRT \\ 
1992.32 & 0.90 & 5.64 & 5.43 & 5.09 &      &      & TCS \\ 
1992.40 & 0.90 & 5.57 & 5.37 & 5.03 & 4.80 &      & TCS \\ 
1992.55 & 0.92 & 5.61 & 5.40 & 5.04 & 4.84 &      & TCS \\ 
1992.61 & 0.93 & 5.58 & 5.38 & 5.05 & 4.9: &      & TCS \\ 
1992.63 & 0.93 & 5.63 & 5.44 & 5.12 & 4.80 & 4.48 & UKIRT \\ 
1992.67 & 0.94 & 5.61 & 5.41 & 5.08 & 4.9: &      & TCS \\ 
1992.82 & 0.96 & 5.63 & 5.39 & 5.04 &      &      & TCS \\ 
1992.88 & 0.96 & 5.62 & 5.40 & 5.06 & 4.83 &      & TCS \\     
1993.03 & 0.98 & 5.58 & 5.36 & 5.00 & 4.94 &      & TCS* \\ 
1993.32 & 0.02 & 5.38 & 4.55 & 3.54 & 2.25 & 1.91 & UKIRT \\ 
1993.34 & 0.02 &      & 4.67 & 3.54 &      &      & Calgary  \\ 
1993.36 & 0.03 &      & 4.53 & 3.46 &      &      & Calgary* \\ 
1993.40 & 0.03 & 5.40 & 4.71 & 3.66 & 2.27 & 1.97 & UKIRT \\ 
1993.47 & 0.04 & 5.49 & 4.86 & 3.89 & 2.43 & 2.1: & UKIRT \\ 
1993.62 & 0.06 & 5.50 & 4.95 & 4.01 & 2.41 &      & TCS \\ 
1993.66 & 0.06 & 5.56 & 5.03 & 4.16 & 2.48 &      & TCS \\ 
1994.13 & 0.12 & 5.72 & 5.22 & 4.60 & 3.11 &      & TCS \\ 
1994.20 & 0.13 & 5.63 & 5.26 & 4.52 & 2.9: &      & TCS \\ 
1994.28 & 0.14 & 5.74 & 5.26 & 4.60 & 3.05 &      & TCS \\ 
1994.40 & 0.16 & 5.63 & 5.28 & 4.64 & 2.99 &      & TCS \\ 
1994.70 & 0.19 & 5.57 & 5.28 & 4.72 & 3.6: &      & TCS \\ 
1994.93 & 0.22 & 5.61 & 5.38 & 4.87 & 3.96 &      & TCS \\ 
1995.33 & 0.27 & 5.59 & 5.36 & 4.95 & 4.16 &      & TCS \\ 
1995.57 & 0.30 & 5.57 & 5.38 & 4.96 &      &      & TCS \\
1995.60 & 0.31 & 5.56 & 5.34 & 4.94 &      &      & TCS \\   
1995.62 & 0.31 & 5.64 & 5.36 & 5.01 & 4.20 &      & TCS \\ 
1995.64 & 0.31 & 5.60 & 5.36 & 4.98 &      &      & TCS \\               
1995.89 & 0.34 & 5.52 & 5.35 & 4.98 &      &      & TCS \\                
1996.32 & 0.40 & 5.61 & 5.42 & 5.06 &      &      & TCS \\         
1996.39 & 0.41 & 5.62 & 5.39 & 5.03 &      &      & TCS \\             
1996.51 & 0.42 & 5.65 & 5.46 & 5.02 &      &      & TCS \\  
1996.59 & 0.43 & 5.59 & 5.36 & 5.02 &      &      & TCS* \\                
1996.61 & 0.43 & 5.62 & 5.40 & 5.02 & 4.39 &      & TCS \\ 
1996.74 & 0.45 & 5.58 & 5.50 & 4.90 & 4.55 &      & TIRGO \\ 
1997.58 & 0.56 & 5.62 & 5.40 & 5.05 & 4.7: &      & TCS \\ 
1997.63 & 0.56 & 5.67 & 5.46 & 5.09 & 4.56 &      & TCS \\
1997.64 & 0.56 & 5.63 & 5.43 & 5.09 & 4.73 &      & TCS \\ 
1997.65 & 0.57 & 5.66 & 5.42 & 5.08 & 4.57 &      & TCS* \\ 
1998.72 & 0.70 & 5.63 & 5.42 & 5.06 &      &      & TCS \\   
1999.55 & 0.80 & 5.60 & 5.38 & 5.03 &      &      & TCS \\
2001.25 & 0.02 & 5.1: & 4.3: & 3.2: & 2.1: &      & TIRGO \\  
\hline
\end{tabular}
\label{TnIRP}
\end{table}  

\begin{table}
\caption{New mid-infrared photometry of WR\,140}
\begin{tabular}{lccccc}
\hline
Date    &$\phi$&   [8.75]      &     [12.5]      & [19.5]      & Tel \\
\hline
1990.66 & 0.69 & 3.40$\pm$0.05 &  3.01$\pm$0.10  & 2.0$\pm$0.4 & UKIRT \\ 
1992.63 & 0.93 & 3.91$\pm$0.06 &  3.13$\pm$0.11  & 2.6$\pm$0.4 & UKIRT \\ 
1993.50 & 0.04 & 1.68$\pm$0.06 &  1.60$\pm$0.08  & 1.8$\pm$0.2 & UKIRT \\ 
1994.32 & 0.15 & 1.64$\pm$0.10 &  1.65$\pm$0.05  &             & UKIRT \\ 
1994.36 & 0.15 &               &                 & 1.6$\pm$0.1 & UKIRT \\ 
2000.69 & 0.95 & 4.01$\pm$0.20 &  3.26$\pm$0.30  &             & SPM \\ 
\hline
\end{tabular}
\label{TmIRP}
\end{table}  

\section{Observations}\label{SObs}      
\subsection{Photometry}

The near-IR photometry (Table \ref{TnIRP}) was observed with a variety of 
telescopes: the United Kingdom Infrared Telescope (UKIRT), the Carlos 
S\'{a}nchez Telescope (TCS) of the Instituto Astrof\'{\i}sica de Canarias, 
the Telescopio InfraRosso del Gornergrat (TIRGO), the 2.1-m telescope of the 
Observatorio Astr\'{o}nomico Nacional at San Pedro M\'{a}rtir (SPM) and the 
University of Calgary 2-m Infrared Telescope (Calgary). Most of the observations 
were fitted in the gaps of other programmes, and the photometry is not expected 
to be as homogeneous as that from a single instrument or campaign. 
The uncertainties of the $J$, $H$ and $K$ magnitudes are typically 0.04 mag., 
while those of $L^{\prime }$ ($\lambda = 3.8\umu$m, $\Delta\lambda = 0.7\umu$m) 
and $nbM$ ($\lambda = 4.63\umu$m, $\Delta\lambda = 0.17\umu$m) are less than 
0.1 mag except where marked `:'. The $nbM$ filter differs slightly from the 
$M$ filter used for Paper~1 but examination of the two datasets showed no 
systematic difference between $M$ and $nbM$ magnitudes, so we combined them.
Comparison of the photometric standards used in common by UKIRT and the TCS 
revealed differences of 0.04, 0.03 and 0.04 mag in $J$, $H$ and $K$ respectively, 
and these shifts were added to the TCS magnitudes in Table \ref{TnIRP} for 
subsequent use. We made no adjustments to the other datasets. 
The final observation, although taken under hazy conditions, signalled the 
2001 dust formation episode at $\phi = 0.02$.

The mid-IR data (Table \ref{TmIRP}) were observed with the bolometer photometer, 
UKT8, on UKIRT, or the Dual Infrared Camera, CID (Salas, Cruz-Gonz\'alez \& 
Tapia 2006), on the SPM 2.1-m telescope. The passbands of the intermediate 
bandwidth filters used for the CID observations differ slightly from the 
corresponding filters in UKT8, but the colour terms are expected to be 
smaller than the photometric uncertainties.

\subsection{Imaging}

\begin{figure}                                      
\includegraphics[clip,width=8cm]{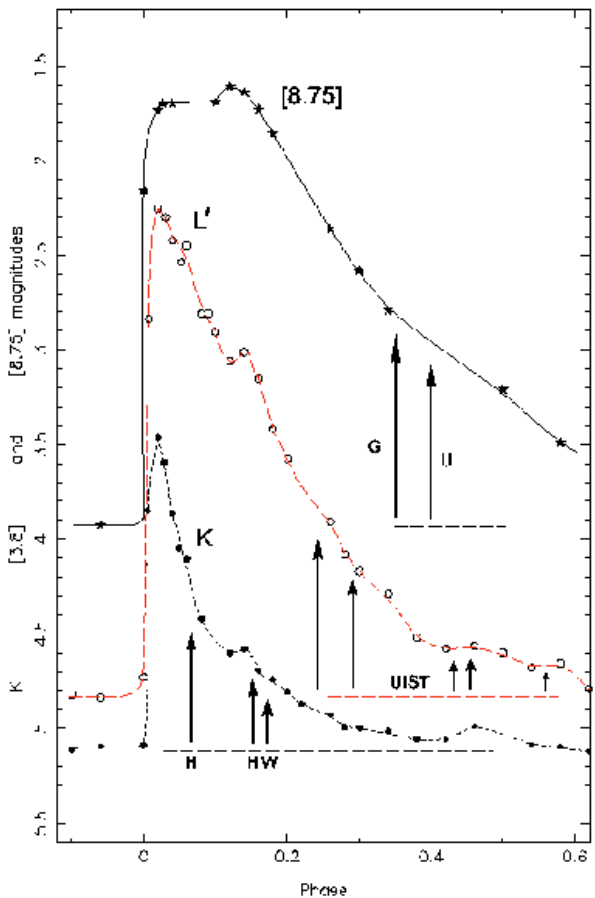}
\caption{Infrared light curves of WR\,140 at three wavelengths 
(8.75, 3.8 and 2.2 $\umu$m) showing the evolution of the dust emission. 
The phases of the imaging observations are marked on appropriate 
light curves. On the $K$-band curve, the phases of our PHARO 
$K$-band images are marked `H', while that of our INGRID 2.27-$\umu$m 
observation is marked `W' (the third PHARO observation at almost the same 
phase is omitted for clarity). The horizontal dashed line fitted to the 
pre-eruption photometry is assumed to give the underlying stellar flux, 
so that the contrast between dusty system and stellar fluxes at the times 
of the imaging observations are readily seen. Similarly, the phases 
of our UIST $nbL^{\prime}$, 3.99-$\umu$m and $M^{\prime}$ observations are 
marked on the $L^{\prime}$ (3.8-$\umu$m) light curve, and the Gemini North 
and UKIRT Michelle observations are marked `G' and `U' on the 8.75-$\umu$m 
light curve, with dashed lines again marking pre-eruption flux levels 
assumed indicative of the stellar fluxes.}
\label{Partlc}
\end{figure}

The imaging observations (Table \ref{Timages}) were made with four different 
instruments: the Palomar High Angular Resolution Observer (PHARO) 
(Hayward et al. 2001) and AO system on the 5-m Hale telescope;  
the Isaac Newton Group Red Imaging Device (INGRID) (Packham et al. 2003) 
and NAOMI AO system on the 4.2-m William Herschel Telescope (WHT);  
the UKIRT 1--5 micron Imager Spectrometer (UIST) (Ramsay Howat et al. 2004) 
on the 3.8-m United Kingdom Infrared Telescope (UKIRT) and the Mid-infrared 
Imager/spectrometer (Michelle) (Glasse, Atad-Ettedgui \& Harris 1997) on 
both UKIRT and the Gemini North telescope. 

The PHARO observations were made on 2001 September 5, 2002 April 22 and 
2002 July 25. The shortest frame time (1817 ms) available to the PHARO 
system required use of a 1 per cent neutral-density filter in conjunction 
with the $K$ filter. The pixel size was 25 mas and the position of WR\,140 
on the array was dithered between images. Prior to observation, the 
adaptive optics system was set up using the point-spread function (PSF) 
standards, HD\,203112 and HD\,203856. 
Observations of WR\,140 and the PSF standards were interspersed so that 
measurements were made at very similar air masses. Images of the PSF 
standard, HD\,203856, were used in further processing of each set of images 
using {\sc mem2d} routines in IRAF. Checks using images of HD\,203112 for 
the PSF yielded consistent deconvolved images.

\begin{figure*}                                             
\includegraphics[clip,width=17.5cm]{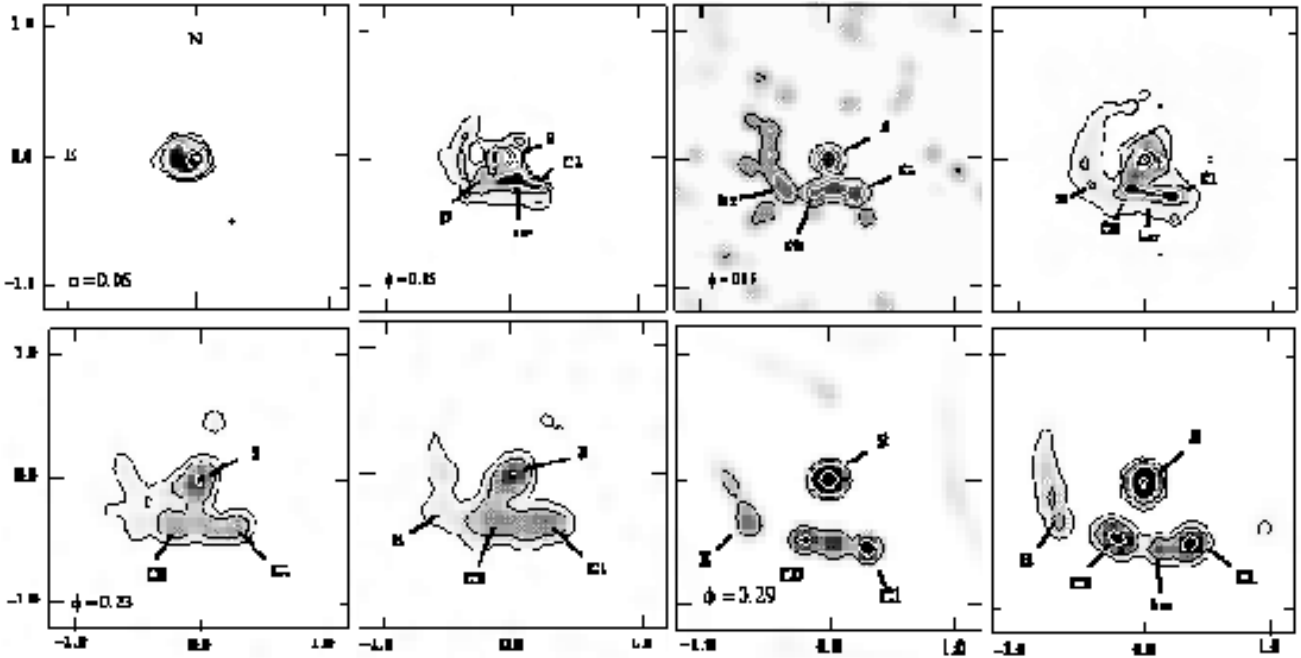}
\caption{Images presented at the same scale (2.4 arcsec square) and orientation 
to show evolution of the dust emission from $\phi$ = 0.06 to 0.29. Superimposed on 
the grey-scale images are contours drawn at the same, equal logarithmic flux 
intervals (0.8 dex, 2 mag.). The images are centred on the star (`S') and the 
positions of the dust emission features `C1', etc., are given in Table \ref{Kpos}.
Top (left--right): two-micron images observed on 2001 September 5 ($\phi = 0.06$) 
and 2002 April 22 ($\phi = 0.15$) with the PHARO/AO system on the Hale telescope. 
The third and fourth images were observed within three weeks of each other at 
$\phi = 0.18$, on 2002 July 4 with the NAOMI/INGRID system on the WHT and on 
July 22 with the PHARO/AO system. The PSF was measured from the comparison 
star observations to be 0.19 arcsec (FWHM).
Bottom (left--right) Pairs of images through the $nbL^{\prime}$ and [3.99] filters 
observed with UIST on UKIRT on 2002 November 20 ($\phi = 0.23$) and 2003 June 4 
($\phi = 0.29$). Owing to the longer wavelengths of observations, their resolution 
is about half that of the two-micron images, with typical PSFs of 0.36 arcsec 
(FWHM) measured from the comparison stars.}
\label{FPharo}
\end{figure*}

The INGRID observation was made on 2002 July 4 through the narrow-band 
$K$-cont filter ($\lambda = 2.27\umu$m, $\Delta\lambda = 0.03\umu$m). 
One hundred 3-sec integrations of WR\,140, dithered in a five-point 
pattern on the chip, were combined using Starlink {\sc ccdpack} routines. 
These data were taken in two runs, separated by sets of 10-s integrations 
of the PSF star HD\,203856. Initially, the WR\,140 data in the two runs 
were reduced separately to allow comparison of the two combined images. 
Both showed very similar structures, giving confidence that they were real. 
The images were then reconstructed with the {\sc mem2d} maximum entropy 
routine using the combined images of HD\,203856 for the PSF. 
The reconstructed image is shown in Fig. \ref{FPharo}. The central 
wavelengths of the $K$ and $K$-cont filters are close enough for us 
to treat the PHARO and INGRID images together, and we will refer to them 
as `two-micron images'.

From the $K$-band light curve (Fig.\,\ref{Partlc}), it can be seen that the 
contrast, the difference between the total (dust + star) and stellar 
emission (assumed to be equal to the pre-outburst emission), fell from 0.85 
to 0.35 mag during our sequence of two-micron imaging observations.

\begin{table}
\caption{Log of imaging observations of WR\,140 including wavelengths and 
pixel sizes.}
\begin{tabular}{lcccr}
\hline
Telescope:      & Date    & Phase &$\lambda$obs & pixel  \\
Instrument      &         &       &($\umu$m)  & (mas)    \\
\hline
Hale: PHARO      & 2001.68 & 0.06 & 2.2   & 25.0  \\
Hale: PHARO      & 2002.31 & 0.15 & 2.2   & 25.0  \\
WHT: INGRID      & 2002.51 & 0.18 & 2.27  & 38.0  \\ 
Hale: PHARO      & 2002.56 & 0.18 & 2.2   & 25.0  \\
UKIRT: UIST      & 2002.89 & 0.23 & 3.6   & 60.6  \\ 
UKIRT: UIST      & 2002.89 & 0.23 & 3.99  & 61.5  \\
UKIRT: UIST      & 2003.42 & 0.29 & 3.6   & 60.6  \\
UKIRT: UIST      & 2003.42 & 0.29 & 3.99  & 61.5  \\
Gemini: Michelle & 2003.86 & 0.35 & 7.9   & 100.5 \\
Gemini: Michelle & 2003.86 & 0.35 & 12.5  & 100.5 \\
Gemini: Michelle & 2003.96 & 0.36 & 7.9   & 100.5 \\
Gemini: Michelle & 2003.96 & 0.36 & 12.5  & 100.5 \\
Gemini: Michelle & 2003.96 & 0.36 & 18.5  & 100.5 \\
UKIRT: Michelle  & 2004.25 & 0.40 & 10.5  & 210   \\  
UKIRT: UIST      & 2004.49 & 0.43 & 3.6   & 60.6  \\
UKIRT: UIST      & 2004.49 & 0.43 & 3.99  & 61.1 \\
UKIRT: UIST      & 2004.49 & 0.43 & 4.68  & 63.5 \\
UKIRT: UIST      & 2004.71 & 0.46 & 3.99  & 61.1 \\
UKIRT: UIST      & 2004.71 & 0.46 & 4.68  & 63.5 \\
UKIRT: UIST      & 2005.52 & 0.56 & 3.99  & 61.5 \\
UKIRT: UIST      & 2005.52 & 0.56 & 4.68  & 63.6 \\
\hline
\end{tabular}  
\label{Timages}
\end{table}

The UIST observations were all made in the UKIRT Service Observing programme.  
Image stabilization was provided by an active tip-tilt secondary. The 
observations were taken by nodding the telescope to four points on the 
array, subtracting adjacent frames and adding the pairs. The PSF standards 
were observed on the same regions of the array as WR\,140 and with the same 
jitter pattern. Individual integration times were typically 4~s in $nbL^{\prime}$ 
and [3.99] and 0.4~s in $M^{\prime}$, repeated to give total integration times 
of, typically, 400~s on WR\,140.
Preliminary reduction of the data was done using {\sc oracdr}, the pipeline data 
reduction at the telescope.  Final reduction, involving matching the centroids, 
averaging the frames, and {\sc mem2d} reconstruction was then done using the 
Starlink packages {\sc kappa} and {\sc ccdpack}.

During the campaign, we observed through progressively longer wavelength 
filters as the dust cooled and the contrast at the shorter wavelengths fell. 
On 2002 November 20 and 2003 June 4 we used the narrow-band $nbL^{\prime}$ 
($\lambda$ = 3.6 $\umu$m) and [3.99] ($\lambda$ = 3.99 $\umu$m) filters; 
on 2004 June 27, we added the $M^{\prime}$ ($\lambda$ = 4.68 $\umu$m) filter, 
which is similar to the $nbM$ filter used for the photometry. As the dust 
cooled further, we dropped the $nbL^{\prime}$ filter for the 2004 September 18 
and 2005 July 21 observations.  
The longer wavelengths and slightly smaller aperture of the telescope meant 
that these images had about half the spatial resolution of our two-micron 
images observed in 2002. From the comparison stars, we measured PSFs 
of 0.36 arcsec (FWHM) at 3.99 $\umu$m, compared with 0.19 arcsec (FWHM) on 
the 2.27-$\umu$m INGRID images. 

The PSF standard for UIST observations in 2004--5 was the UKIRT photometric 
standard HD\,201941, allowing calibration of the images. This yielded  
$M^{\prime}$ = 4.11, 4.16 and 4.34 at phases 0.43, 0.46 and 0.56 
respectively, in excellent agreement with the $M/nbM$ light curve. 
 
Images in the mid-IR were observed using Michelle on Gemini North on 2003 
November 9 and December 15. The data were acquired in a regular chop-nod 
pattern, with an on-chip 15-arcsec E--W chop throw and N--S nodding. 
For combination of images coming from multiple exposures, we used only the 
chopped images taken under active guidance with a peripheral wavefront 
sensor. 
The images have resolutions (0.50 arcsec at 7.9$\umu$m and 0.59 arcsec at 
12.5$\umu$m, from observations of the standard star PSFs) somewhat lower than 
those of the UIST images because the longer wavelengths outweigh the advantages 
of the larger aperture of Gemini.

Another mid-IR observation with Michelle, this time on UKIRT, was made on 
2004 March 31 in the UKIRT Service Observing Programme. The secondary was 
chopped 15 arcsec N--S and the telescope nodded 15 arcsec E--W.  
The final image was formed by matching the centroids of the four stellar 
images and averaging them. The object and the standard (BS 7525) were observed 
on the same region of the array with the same nod and chop offsets and  
exposure times.
Preliminary reduction of these data used {\sc oracdr}, and final reduction, 
involving matching the centroids, averaging the frames, and {\sc mem2d} 
reconstruction used the Starlink packages {\sc kappa} and {\sc ccdpack}. 
The derived magnitude, $N = 2.9\pm0.2$, of the (star+dust) system at phase 
0.40 is consistent with the [8.75] and [12.5]-band light curves.

\begin{figure*}                                          
\includegraphics[clip,width=17.5cm]{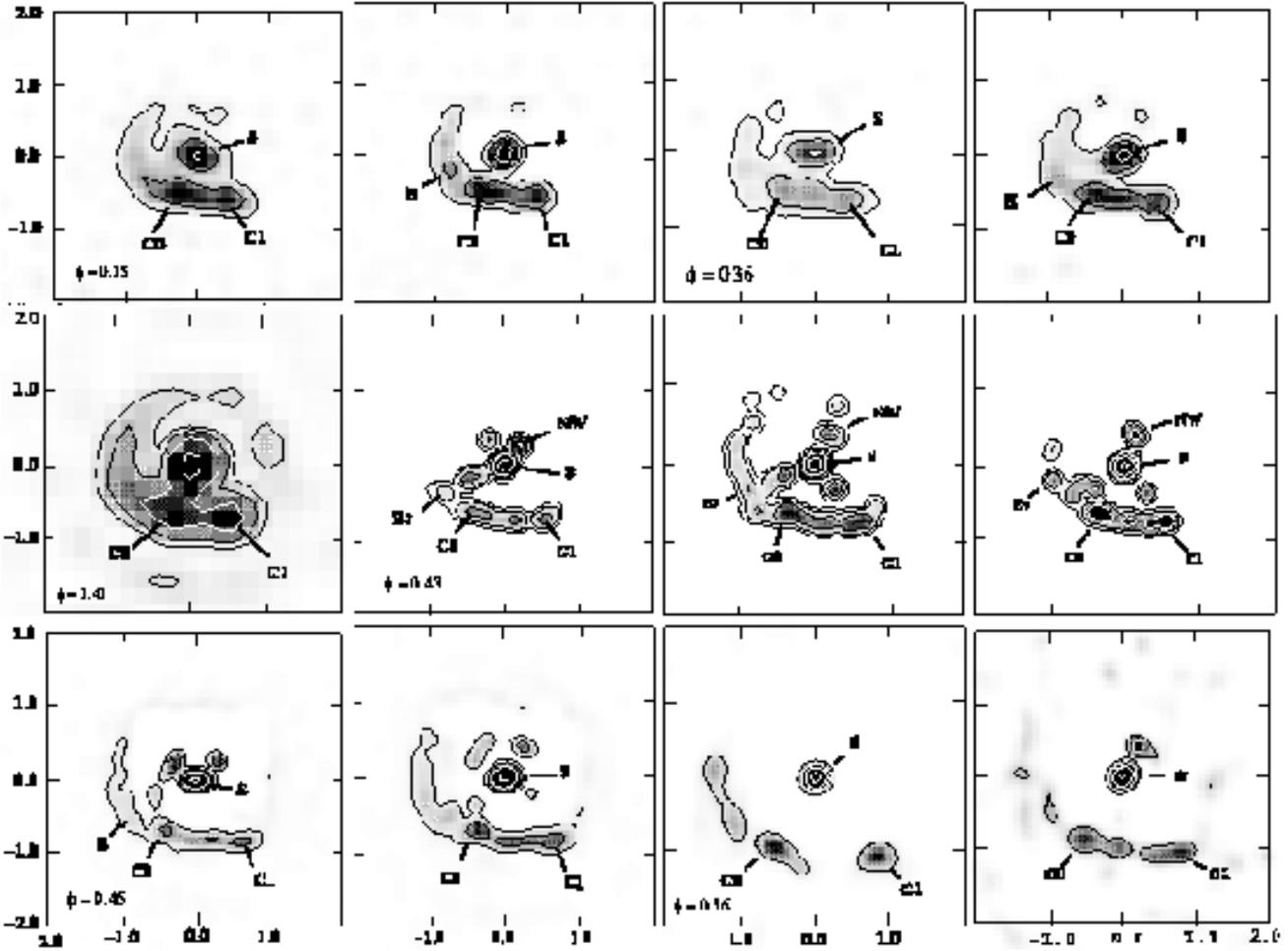}
\caption{Images presented at a larger scale (4 arcsec square) to show later 
evolution of the dust emission ($\phi = 0.35 - 0.56$); the orientations 
and contour intervals are as in Fig.\,\ref{FPharo}. 
Top row (l--r): Pairs of images observed with Michelle on Gemini North through 
the narrow-band 7.9-$\umu$m and 12.5-$\umu$m filters on 2003 November 9 ($\phi = 0.35$) 
and December 15 ($\phi = 0.36$). The PSFs of the comparison stars were measured to 
be 0.50 arcsec at 7.9$\umu$m and 0.59 arcsec at 12.5$\umu$m, slightly larger than 
those of the UKIRT/UIST $nbL^{\prime}$ and [3.99] images as the longer wavelengths 
are partly compensated for by the larger aperture of Gemini.
Middle row (l--r): Image at 10.5$\umu$m observed with Michelle on UKIRT on 2004 
March 31 ($\phi = 0.40$), which has the lowest resolution (PSF 0.8 arcsec FWHM) on 
account of the wavelength and telescope aperture, followed by three images observed 
with UIST on UKIRT on 2004 June 27 ($\phi = 0.43$) through the $nbL^{\prime}$, 
[3.99] and $M^{\prime}$ filters.  
We are not sure we have measured `E' in these and find compact features considered 
to be an artefacts (e.g. `NW', see text) nearer the star. These appear in the 
subsequent images below, and do not share the movement of the dust features away 
from the star. 
Bottom row (l--r): Pairs of images observed with UKIRT/UIST through the [3.99] and 
$M^{\prime}$ filters on 2004 September 25 ($\phi = 0.46$) and 2005 July 21 
($\phi = 0.56$).}
\label{FUIST}
\end{figure*}

\begin{figure*}                                        
\hfill
\includegraphics[clip,width=15.4cm]{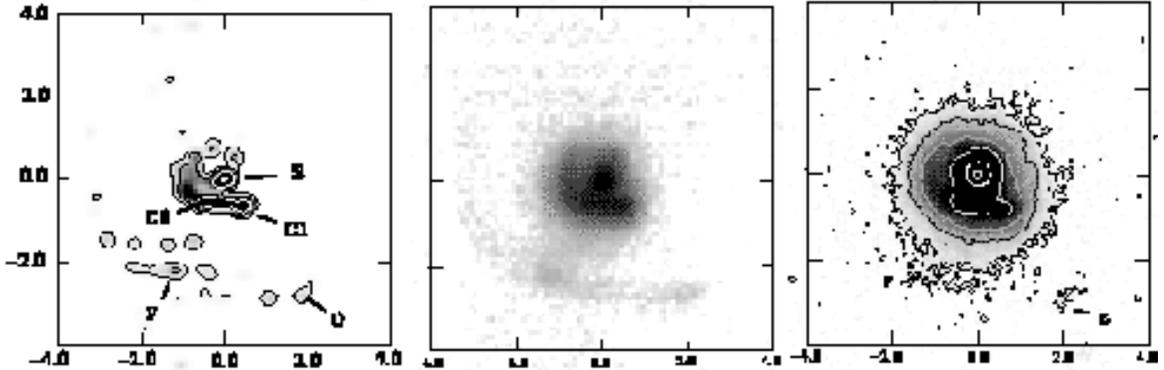} 
\hfill
\caption{Images at a larger scale (8 arcsec square) to show the faint emission 
features believed to be the remnants of the dust formed in the 1993 episode, 
including `F' and `G', which have P.A.s very close to those of `C0' and `C1' 
and distances consistent with their being one period older (Section \ref{SImages}).
Left: the 2003 December 12.5-$\umu$m image with lowest contour level (3.7$\sigma$ sky) 
well below that (20 $\sigma$ sky) in the corresponding image in Fig.\,\ref{FUIST}. 
Centre: the November and December 12.5-$\umu$m images combined, and plotted with 
an intensity scale chosen to emphasise the faint emission features. 
Right: The unreconstructed 2004 September 25 $M^{\prime}$ image at an intensity scale 
chosen to bring out the fainter emission, including `F' on the outskirts of the central 
emission the equivalent of `C0' formed in the 1993 dust-formation episode.}
\label{FGem}
\end{figure*}


\begin{table*}
\caption{Position angles (P.A., deg. East of North) and projected distances 
 ($\xi$, mas), relative to the central star, of dust `knots' identified in the IR images. 
 The positions from the Keck observations come from MTD except those 
 marked (*), which were made by ourselves from their 2.21-$\umu$m image or the 
 3.08-$\umu$m image in Tuthill et al. (2003).}
\begin{tabular}{lcccrrrrrrrrrr}
\hline
Tel: Inst    & Date & Phase & $\lambda$obs & \multicolumn{2}{|c|}{knot C1} & \multicolumn{2}{|c|}{knot C0} & \multicolumn{2}{|c|}{knot D} & \multicolumn{2}{|c|}{knot E}\\
             &       &      &($\umu$m)    & P.A. & $\xi$& P.A. & $\xi$& P.A. & $\xi$& P.A. & $\xi$\\
\hline 
Keck (MTD)   & 2001.45 & 0.04 & 2.21   & 231  &   54 &      &      & 134  &   77 & 113  &  110 \\
Keck (MTD)   & 2001.58 & 0.06 & 2.21   & 227  &   76 & *154 &  *64 & 136  &  125 & 113  &  172 \\
Keck (Tuthill et al.) & 2001.58 & 0.06 & 3.08   & *229 &  *91 &      &      & *133 &  *99 & *114 & *175 \\
Hale: PHARO  & 2002.31 & 0.15 & 2.2    & 216  &  278 & 161  &  204 &      &      &      &      \\
WHT: INGRID  & 2002.51 & 0.18 & 2.27   & 219  &  343 & 157  &  290 & 124  &  419 &      &      \\
Hale: PHARO  & 2002.56 & 0.18 & 2.2    & 217  &  352 & 157  &  261 &      &      & 115  &  467 \\
UKIRT: UIST  & 2002.89 & 0.23 & 3.6    & 217  &  498 & 163  &  330 &      &      &      &      \\
UKIRT: UIST  & 2002.89 & 0.23 & 3.99   & 216  &  460 & 168  &  360 &      &      & 115  &  577 \\
UKIRT: UIST  & 2003.42 & 0.29 & 3.6    & 209  &  626 & 158  &  526 &      &      & 118  &  729 \\
UKIRT: UIST  & 2003.42 & 0.29 & 3.99   & 217  &  601 & 156  &  491 &      &      & 119  &  759 \\
Gemini: Michelle & 2003.86 & 0.35 & 7.9  & 214  &  714 & 153  &  553 &      &      & 112  &  812 \\
Gemini: Michelle & 2003.86 & 0.35 & 12.5 & 213  &  736 & 157  &  603 &      &      & 107  &  842 \\
Gemini: Michelle & 2003.96 & 0.36 & 7.9  & 214  &  755 & 141  &  654 &      &      & 102  &  976 \\
Gemini: Michelle & 2003.96 & 0.36 & 12.5 & 213  &  780 & 145  &  632 &      &      & 110  &  960 \\
UKIRT: Michelle & 2004.25 & 0.40 & 10.5  & 215  &  874 & 162  &  684 &      &      &      &      \\
UKIRT: UIST  & 2004.49 & 0.43 & 3.6    & 216  &  926 & 148  &  766 &      &      &      &      \\
UKIRT: UIST  & 2004.49 & 0.43 & 3.99   & 219  &  986 & 149  &  749 &      &      &      &      \\
UKIRT: UIST  & 2004.49 & 0.43 & 4.68   & 218  &  966 & 157  &  727 &      &      &      &      \\
UKIRT: UIST  & 2004.71 & 0.46 & 3.99   & 216  & 1055 & 149  &  781 &      &      &      &      \\
UKIRT: UIST  & 2004.71 & 0.46 & 4.68   & 218  & 1072 & 154  &  778 &      &      &      &      \\
UKIRT: UIST  & 2005.52 & 0.56 & 3.99   & 218  & 1363 & 149  & 1079 &      &      &      &      \\
UKIRT: UIST  & 2005.52 & 0.56 & 4.68   & 218  & 1326 & 150  & 1013 &      &      &      &      \\ 

\hline
\end{tabular}
\label{Kpos}
\end{table*}

\section{Evolution of the dust emission and expansion of the cloud} 

\label{SEvol}
\subsection{Photometry and properties of the dust grains}
\label{SPhot}

The new data strengthen the determination of the pre-eruption SED, for which we 
adopt: $H$ = 5.43, $K$ = 5.12, $L^{\prime}$ = 4.82, $M/nbM$ = 4.53, [8.75] = 3.95, 
[12.5] = 3.2 and [19.5] = 2.8. They also define the stellar SED, i.e. neglecting 
any contribution from dust made in the 1993 and earlier eruptions (see below). 
These are used as a baseline for modelling the post-eruption SEDs 
and in Section \ref{SPimages} for calibrating the images, assuming the stellar 
flux not to vary over the period of the imaging observations -- a view 
supported by the optical monitoring (MM03), which shows long-term stability 
between periastron passages. 

Comparison of the new data with those in Paper I and Williams et al. (1978) shows 
that the behaviour followed that observed from previous periastron passages and 
allows re-examination of the photometric period. Lafler-Kinman period searches 
on the data in the fading branches of the $H$, $K$ and $L^\prime$ light curves 
give periods of $2905\pm8$, $2900\pm4$ and $2905\pm10$~d. respectively, slightly 
longer than the $2900\pm10$~d. found in Paper I and the RV period 
($2899\pm1.3$~d., MM03). We adopt a weighted mean period of 2900 d. 
Unfortunately, we were not able to observe the much steeper rising branches 
of the light curve in either 1993 or 2001 which, with the 1985 rise, would 
give a significantly more precise period. 

\begin{figure}                                      
\includegraphics[clip,width=7cm]{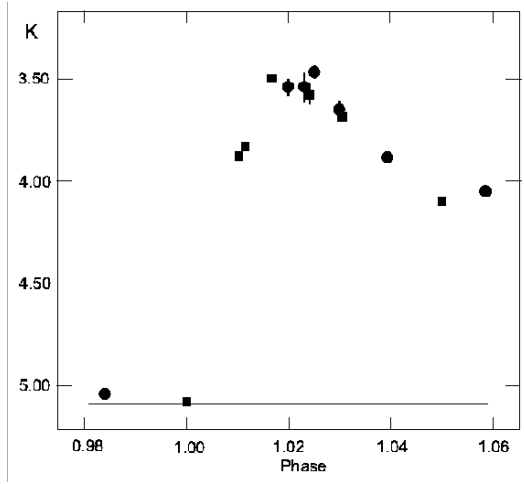}
\caption{Light curve of WR\,140 near periastron in $K$ with observations 
from the 1985 (squares) and 1993 (circles) maxima. The horizontal line 
represents the level of the stellar wind emission.}
\label{Kmax}
\end{figure}

Using this period and MM03's T$_{\rmn 0}$, we produced light curves gridded 
to intervals of 0.02P (0.01P for $0 < \phi < 0.1$) at three representative 
wavelengths (Fig.\,\ref{Partlc}) to illustrate the level of dust emission 
relative to that of the star at the phases of the imaging observations, and 
summarise the evolution of the dust emission. To show the beginning and 
duration of the condensation of new grains, we show (Fig.\,\ref{Kmax}) the 
$K$-band light curve near periastron based on individual magnitudes observed 
during the 1985 and 1993 events. Phased light 
curves in more filters, albeit based on fewer data, were given in Paper~1.

The light curves track the evolution of the dust emission integrated over 
all the individual features comprising different masses of dust at different 
temperatures. An average dust temperature, $\langle T_{\rmn{g}} \rangle$, 
can be determined for the dust at each phase by fitting the multi-wavelength 
photometry with optically thin emission by carbon grains after correcting 
for interstellar reddening and subtracting the stellar SED, as in Paper~1. 
Here we adopted $A_{\rmn{V}} = 2.9$ (Morris et al. 1993) and absorption 
coefficients of the `ACAR' amorphous carbon grains prepared in an inert 
(Argon) atmosphere in the laboratory by Colangeli et al. (1995). 
Cross-sections were calculated 
using the optical properties for this sample given by Zubko et al. (1996). 
Model SEDs calculated using the `ACAR' data gave a better fit to broad 
features in the 5.5--6.5 $\umu$m {\em ISO} spectra of dust-making WC8--10 
stars (Williams, van der Hucht \& Morris, 1998) than those based on the 
`ACH2' grains produced by Colangeli et al. in a hydrogen atmosphere --
as one might expect for grains formed in hydrogen-poor WC stellar winds. 
The temperature, $T_{\rmn{g}}$, of a grain is determined primarily by its 
radiative equilibrium in the stellar radiation field at distances, 
$r_{\rmn{O}}$ and $r_{\rmn{W}}$ from the O5 and WC7 stars: 
\[
4\overline{Q}_{\rmn{a}}(a,T_{\rmn{g}}) T_{\rmn{g}}^4 =
\frac{\overline{Q}_{\rmn{a}}(a,T_{\rmn{O}})T_{\rmn{O} }^4}{(r_{\rmn{O}}/R_{\rmn{O} })^2}
+ \frac{\overline{Q}_{\rmn{a}}(a,T_{\rmn{W}})T_{\rmn{W} }^4}{(r_{\rmn{W}}/R_{\rmn{W} })^2}, 
\]
\noindent where $\overline{Q}_{\rmn{a}}(a,T)$ are the Planck mean absorption 
cross-sections appropriate to the grain or stellar temperature for grains 
of radius $a$, and the factor $4$ is for spherical grains. The ACAR grains 
have $\overline{Q}_{\rmn{a}}(a,T_{\rmn{g}}) \propto T_{\rmn{g}}^{1.2}$ in 
the relevant temperature range, so the radiative-equilibrium grain temperature 
falls off with distance as $T_{\rmn{g}} \propto r^{-0.38}$ for constant grain 
size and stellar luminosity. We assume the distance between the two stars is 
much smaller than that of the dust to either, and use $r$ for 
$r_{\rmn{O}}$ and $r_{\rmn{W}}$. We can derive distances corresponding to the 
average temperatures, allowing us to track the movement of the dust away from 
the stars, especially in the early phases when the dust emission has not been 
imaged. The dust inherits the velocity of the compressed wind from which it 
condenses and will then be accelerated by radiation pressure so, in the absence 
of any other effect, newly made dust will cool as it moves from the stars, and 
its emission will fade. 

Fits to the $H$ to [19.5] and $H$ to [12.5] data at phases 0.01 and 0.02 gave 
$\langle T_{\rmn{g}} \rangle \simeq 1100$K and dust masses 
$M_{\rmn{g}} = 2\times 10^{-8}$ and 3$\times 10^{-8} M_{\odot}$. The approximate 
constancy of $\langle T_{\rmn{g}} \rangle$ and therefore the radiative-equilibrium 
$r$ at these phases while the dust mass rose can be interpreted as continued 
condensation of new dust at a fixed distance from the stars. This is the 
nucleation radius, the closest to the stars that grains can survive sublimation. 
At $\phi = 0.00$, we have only $JHKL^\prime$ data and the fit to them gives 
$\langle T_{\rmn{g}} \rangle \simeq 1200$~K and 
$M_{\rmn{g}} = 2\times 10^{-10} M_{\odot}$ or, fixing the temperature at 1100~K, 
$M_{\rmn{g}} = 3\times 10^{-10} M_{\odot}$, 
so it is evident that dust formation had just begun at periastron. 
This is supported by the individual $JHK$ observations, which show 
(e.g. Fig.\,\ref{Kmax}) no significant excess over wind emission at 
phases $\phi$ = 0.844 and 0.999 and a broad maximum around $\phi = 0.025$. 
By $\phi = 0.03$, the $JHKL^\prime$ magnitudes and temperature were falling, 
indicating that no new dust was condensing to replenish the dust carried 
away with the wind. From $\phi = 0.03$ to $\phi = 0.12$, the near-IR 
flux fell while the mid-IR flux rose to maximum. The model fits to the 
SEDs show that the dust mass doubled to 6$\times 10^{-8} M_{\odot}$ while 
$\langle T_{\rmn{g}} \rangle$ fell to 800~K. This suggests that the increase 
of dust mass was caused by the growth of the recently formed grains at their 
equilibrium temperature rather than by the condensation of fresh grains at 
$T_{\rmn{g}} \simeq 1100$K. Zubko (1998) has shown that carbon grains in 
WC9 stellar winds grow to $a \sim$ 100\AA\ by implantation of impinging 
carbon ions as they move through the wind after acceleration by radiation 
pressure. Evidence for larger 
grains made in WR\,140 comes from eclipses observed in the optical light 
curves between phases 0.020 and 0.055, from which MM03 derived a typical 
size of 0.069 $\umu$m (690\AA) for dust grains in our line of sight to the 
star. 

The $K$ and $L^\prime$ light curves show an inflexion at $\phi = 0.14$, 
suggesting a short-lived increase in dust emission. The $UBV$ photometry 
(MM03) does not show brightening at this time, so we ascribe the extra 
emission to an increase of dust formation rather than increased radiative 
heating of the dust. The $H$ photometry does not show any interuption in 
its fading at this phase, suggesting that the brightening at $K$ and 
$L^\prime$ does not come from the condensation of new, hot dust but from  
a temporary increase in grain growth rate. 

After $\phi = 0.14$, the fading continues at all wavelengths and the dust 
cools as expected. The total mass, however, falls from its maximum of 
6.5$\times 10^{-8} M_{\odot}$ to less than 2$\times 10^{-8} M_{\odot}$ 
at phase 0.56. 
This may be an artefact arising from our use of isothermal dust models, 
but a similar effect was observed when modelling the then available IR 
photometry with radially extended dust parcels having an appropriate 
range of $T_{\rmn{g}}$ in Paper 1, so we believe the effect to be real. 
It suggests that, as grains move through the wind, the rate of destruction 
by thermal sputtering eventually overtakes that of growth by implantation 
of carbon ions (cf. Zubko 1998) and grains are destroyed.

The temperatures and dust masses derived here differ from those in Paper~1 
because we used different optical constants for the grains. We note that 
the condensation temperature, 1100~K, found here is lower 
than those found from modelling WC9 SEDs by Zubko (1998) or modelling the 
dust pinwheel about WR\,104 by Harries et al. (2004) using the same `ACAR' 
dust analogue and optical coefficients. This suggests that, either the 
dust formed by WR\,140 has different optical properties from that made by 
the WC9 stars, or that a difference in conditions, e.g. stronger radiation 
field or faster stellar winds, affects the condensation. 

We estimate the nucleation radius, the distance from the stars corresponding 
to the condensation temperature, by modelling the heating of the dust. 
Because the UV-optical spectra of hot stars are not Planckian, we 
calculated model atmosphere analogues of Planck mean cross-sections 
$\overline{Q}_{\rmn{a}}(a,T_{\rmn{O}})$ and 
$\overline{Q}_{\rmn{a}}(a,T_{\rmn{W}})$ using the 35kK WM-Basic O-star 
model fluxes tabulated by Smith, Norris \& Crowther (2002) for the O5 star, 
and their 70kK CMFGEN WC model fluxes for the WC7 component, and absorption 
coefficients for small ACAR grains. 
The temperature chosen for the O-star model comes from the calibration 
of WM-Basic models by Garcia \& Bianca (2004) and our revised type of O5. 
The luminosity and radius of the O5 star were determined by fitting the 
WM-Basic flux to its de-reddened $v$ magnitude and the distance of 1850~pc 
determined by Dougherty et al. (2005), giving $R_{\rmn{O}} = 26R_\odot$ 
and log (L/L$_\odot$) = 5.93. The luminosity is lower than that 
(log (L/L$_\odot$) = 6.18) adopted by Pittard \& Dougherty (2006) and 
close to the typical values (5.6--5.93) given for O5~I--III stars by 
Repolust, Puls \& Herrero (2004). We retained the Pittard \& Dougherty 
luminosity of the WC7 star (log (L/L$_\odot$) = 5.5); it provides about 
one-quarter of the radiation heating the grains. 

The equilibrium distance corresponding to $T_{\rmn{g}} = 1100K$ for small 
(10--50\AA) grains is $r\simeq630$~au. For a distance of 1850 pc 
(Dougherty et al. 2005), this would subtend 340 mas, significantly greater 
than the distances (54--220 mas) at which MTD observed condensations in 
2001 June ($\phi = 0.039$). Larger grains are relatively more efficient 
at cooling, e.g., 0.1-$\mu$m grains would attain $T_{\rmn{g}} = 1100K$ 
at $r\simeq235$~au, but large grains have to start as condensation nuclei, 
so this cannot account for the discrepancy. Nor is it likely to be an 
inclination effect: although the inclination angles of the trajectories 
of these dust clouds are not known, comparison of their projected 
velocities (MTD and Section \ref{SMotion}) with the wind velocities 
suggests low inclinations. 
Nor is it a consequence of the adopted distance to WR\,140: if WR\,140 
were closer, the stars would be less luminous, their radii smaller, the 
radiative equilibrium distance to the dust smaller in proportion, but 
the angular equilibrium distance would be unchanged. The compressed winds 
within which the grains condense may provide some shielding from the 
stellar radiation, but only a difference in dust properties can account 
for most of the discrepancy. 

A comparison of several laboratory analogues of cosmic amorphous carbon grains 
by Andersen, Lodl \& H\"ofner (1999) shows that their absorption efficiencies 
have fairly similar wavelength dependencies ($\kappa \propto \lambda^{-1.1}$ 
in the IR, but absolute values and ratios of ultraviolet-to-IR coefficients 
ranging by a factor of $\sim$ 5. Consequently, while the 
grain temperatures derived from fitting IR SEDs are relatively insensitive to 
the choice of grain sample, the absolute dust masses and the distances derived 
from grain temperatures through equilibrium in hot star radiation {\em do}
depend on the choice of sample\footnote{Planck mean cross-sections of the 
different samples at typical grain temperatures have similar temperature 
dependence, so the $r^{-0.38}$ rate at which $T_{\rmn{g}}$ falls off with 
distance is also fairly insensitive to choice of sample, as are the 
{\em relative} masses of dust determined from SEDs observed at different 
phases.}.

We suggest that the dust formed by WR\,140 has a lower ratio of ultraviolet 
to IR opacities than the `ACAR' sample. For a rough approximation, we introduce 
a scaling factor to the UV opacity scale so as to produce an equilibrium 
temperature at a typical distance (150 au) of a condensation in MTD's 2001 
June image equal to that determined from the SED, 
$\langle T_{\rmn{g}} \rangle = 980$K, at that phase ($\phi = 0.039$), 
using a grain size of 100\AA\ to allow for grain growth. With this scaling, 
the equilibrium distance for small grains would then be a more plausible 125~au.

We continue by applying the same scale factor to calculate the Planck mean 
cross-sections for radiation pressure, $\overline{Q}_{\rmn{pr}}(a,T_{\star})$ 
and hence the acceleration of the grains:
\[
\ddot{r} = \frac{3\sigma}{4acs_{\rmn{g}}}\left(
\frac{\overline{Q}_{\rmn{pr}}(a,T_{\rmn{O}})T_{\rmn{O} }^4}{(r_{\rmn{O}}/R_{\rmn{O} })^2}
+ \frac{\overline{Q}_{\rmn{pr}}(a,T_{\rmn{W}})T_{\rmn{W} }^4}{(r_{\rmn{W}}/R_{\rmn{W} })^2} 
\right).
\]
\noindent Assuming a bulk density $s_{\rmn{g}}$ = 2~g~cm$^{-3}$ for the grain 
material, the acceleration at $r = 125$ au would be about 33 km~s$^{-1}$ per day 
for a 10-\AA\ grain and not much less, 30 km~s$^{-1}$ per day, for a 100-\AA\ grain. 
After only a few days, the grains would have significant drift velocities, $u$, 
relative to the material in which they formed. The principal limitation on the 
acceleration is supersonic drag, which is approximately proportional to $u^2$ and 
the local density, $\rho(r)$. As the grains accelerate, the drag increases until 
it balances the radiation pressure. If the grains are moving through a wind having 
$\rho(r)\propto r^{-2}$, both forces are proportional to $r^{-2}$, and the grains 
will take up a terminal drift velocity relative to the wind. The imaging observations 
allow us to estimate this velocity (Section \ref{SMotion}).

\subsection{The images and expansion of the dust cloud}   
\label{SImages} 

Our images of WR\,140 in Figs \ref{FPharo}--\ref{FGem} do not resolve the binary 
itself, which we will refer to as the star here. To allow easy intercomparison, the 
two-micron images and the first sets of UIST images (Fig.\,\ref{FPharo}) have been 
plotted on a common scale (2.4 arcsec square), and most of the subsequent images 
(Fig.\,\ref{FUIST}) on a larger (4 arcsec square) common scale to accommodate the 
expanded dust cloud.
 
The first PHARO observation (Fig.\,\ref{FPharo}) was made five weeks after the 
2001 July 30 Keck observations by MTD and Tuthill et al. (2003). It shows 
extended emission to the South and East of the central star, consistent with 
the Keck images, but does not resolve the dust emission features. 

By the time of the second PHARO observation in 2002 April, the cloud had expanded 
sufficiently for us to resolve some of the dust features previously observed in the 
Keck images. 
Following MTD, we identified prominent dust-emission peaks and measured their 
positions relative to the star. The position angles (P.A., E of N) and 
projected distances ($\xi$) are given in Table~\ref{Kpos}. At the scale of 
these images, uncertainties of $\pm$1 pixel in the measured positions yield 
uncertainties of 5$\degr$ in P.A. and 30 mas in $\xi$. 
The principal feature apparent in the images here is a `bar' of emission 
(marked `bar' in Fig.\,\ref{FPharo}) to the South of the central star 
(marked `S'), and extending E--W. The `bar' is evident in the  
2.26-$\umu$m Keck image of 2001 July (Tuthill et al. 2003), but only the 
peaks at the two ends are seen in their 3.08-$\umu$m image observed on the 
same date. We label the emission peaks at the E and W ends of the `bar' as 
`C0' and `C1' in all our images. 
Comparison of the morphology of the `bar' structure in the Keck images and 
those presented here suggests a connection between `C1' and their `Feature C' 
but the P.A.s differ by about 12$\degr$, arguing against their being the same 
physical entity in radial expansion. We estimated the P.A. of a concentration 
having a P.A. similar to that of `C1' in the published 2001 July 2.21-$\umu$m 
Keck image and give its position in Table \ref{Kpos}. 

Our April PHARO image shows other structures seen in the Keck images, notably 
the `arm' to the East, which lies further from the star than the `bar', and an 
intermediate concentration which we tentatively identify with MTD's `Feature D' 
on the basis of its P.A. (Table \ref{Kpos}). We measured the $K$-band flux 
of the whole structure using a 0.9-arcsec software aperture and that of the 
star alone with a 0.18-arcsec aperture, and determined a difference of 0.6 mag. 
This difference is consistent with the $K$-band light curve (Fig.\,\ref{Partlc}) 
at the time of observation.

The next two images in Fig.\,\ref{FPharo} come from observations made within 
three weeks of each other (July 4 and July 25) using two different instruments, 
(INGRID and PHARO) and reduced independently using different software suites. 
Assuming there was no significant change in the morphology of the dust emission 
(apart from expansion of the cloud) during this time, 
comparison of the images provides an excellent external check on the veracity 
of the structures revealed in the maximum-entropy reconstructions. Again, the 
most conspicuous feature is the E--W `bar' to the S, at the ends of which we 
have measured the positions of `C0' and `C1'. The P.A.s (Table \ref{Kpos})
agree well with those observed in the April image, and the radial distances 
show continued expansion of the dust structure. We measured the intensity 
profiles of `C1', finding it to be extended E--W, but unresolved N--S, 
consistent with its being a concentration in a thin, elongated structure.
Although we refer to dust emission features as `concentrations', we are 
aware that these are not necessarily physical clumps. The dust is optically 
thin in the IR (Paper~1) and we expect it to form extended, hollow structures 
owing to its origin in a hollow, thin WCR (cf. Section \ref{SModel}), so that 
the brightest emission could be coming from the limb-brightened edges of the 
structures.

Both the INGRID and PHARO images show the `arms' of emission to the East of 
the star, but somewhat differing in form: that in the PHARO image has the 
same angular extent (P.A. 15$\degr$--125$\degr$) as in the Keck images, while 
that in the INGRID image is truncated, perhaps in the image deconvolution. 
We measured an emission peak at a P.A. close to MTD's Feature `E' in the 
PHARO image and tentatively one about $10\degr$ in P.A. from their `Feature D' 
in the INGRID image (Table~\ref{Kpos}). 
We observe extended emission near MTD's `Feature A' at the N ends of the `arms' 
but not their `Feature B' to the NW of the star. 

The UIST observations were made at longer wavelengths, initially 3.6 and 
3.99 $\umu$m, where the contrast between dust and stellar emission was 
expected to be greater (cf. Fig.\,\ref{Partlc}). The 2002--3 UIST images in 
Fig.\,\ref{FPharo} all show the `bar' to the South which, in the 2003 images, 
appears to be resolved into three components, reminiscent of the INGRID 
two-micron image. These structures, however, may be a consequence of the 
reconstruction of faint diffuse emission in the presence of noise, which can 
produce spurious concentrations.
The N--S image profiles of `C1' have half-widths equal to that of the 
star, so the `bar' is unresolved in this direction. These images also show 
the `arm' of emission to the East, more evidently in the 3.99-$\umu$m images, 
in which we identify knot `E' but not much emission to the N of it, let 
alone anything corresponding to MTD's `Feature A' at the N end. 

In the top panel of Fig.\,\ref{FUIST}, we present the mid-infrared images 
observed with Michelle on Gemini North. The structures of the dust clouds 
look very similar to those observed earlier, and we can identify and measure 
most of the same emission features. 
Their positions are included in Table \ref{Kpos}. The P.A.s of `E' show 
some scatter as we measure different points along the `arm' owing to its 
relative faintness and the fact that we are measuring a bright part of 
extended emssion, but the radial distances are more certain 
and show continued uniform expansion of the dust cloud. 

The December 12.5-$\umu$m image shows faint emission further south of the 
cloud which we identify as a remnant of the dust formed at the time of the 
previous (1993) periastron passage. This image is shown at lower scale in 
Fig.\,\ref{FGem}, both with contours as in the other figures and in 
grey-scale only, combined with the November image to bring out the fainter 
emission. In the former, we have marked two features `F' and `G' 
which have position angles ($150\degr$ and $215\degr$) very close to the 
mean P.A.s of knots `C0' ($154\degr$) and `C1' ($216\degr$) from all our 
observations (cf. Table \ref{Kpos}). If `F' and `G' had travelled from 
the star since the 1993 periastron passage, their average proper motions 
over the intervening 10.6 years would have been 235 and 319 mas~y$^{-1}$ 
respectively, in good agreement with the average proper motions of 
`C0' and `C1' ($244\pm9$ and $326\pm7$ mas~y$^{-1}$) determined from our 
2002--5 observations (Table \ref{Tmotion}, Section \ref{SMotion}). 

The continued linear expansion of the dust cloud points to expansion into a 
low density circumstellar environment, essentially the cavity blown by the 
stellar winds of the two components during their main-sequence lifetimes and 
observed as a minimum in H\,{\sc i} emission by Arnal (2001). 
The cavity has a diameter of 110$\arcmin$ and is also seen as a ragged shell 
at {\em IRAS} wavelengths (Marston 1996). Its dynamical age ($>$ 1 million y) 
suggests that it was formed during the earlier O-star phase of the WC7 star.

In 2004, we re-observed WR\,140 in the mid-IR using Michelle, but this time on 
UKIRT. The resolution of the image (Fig.\,\ref{FUIST}) is lower than those of 
the Gemini observations owing to the smaller aperture of the telescope, but 
the principal dust features were still observable and we give the positions 
of `C1' and `C0' in Table \ref{Kpos}. The P.A. of `C0' differs by $8\degr$ 
from the average P.A. ($154\degr$) for this feature --- which equates to half 
a relatively large (210 mas) pixel in this instrument on UKIRT. The P.A. of 
`C1' agrees with the average within a small fraction of a pixel, and the radial 
distances of both features fit the linear expansion (Fig\,\ref{Fmotion}, 
Table \ref{Tmotion}) to within a quarter pixel.

The other three images in the middle panel of Fig.\,\ref{FUIST} were observed 
with UIST on 2004 June 27, when the contrast between stellar and dust emission 
was much lower (cf. the $L^{\prime}$-band light curve in Fig.\,\ref{Partlc}). 
The principal dust feature, the `bar' including `C0' and `C1', is clearly 
present (and seen in the raw images). We also observe compact features 
nearer the star, one to the NW (marked `NW') at all three wavelengths. 
The $M^{\prime}$ UIST images observed in 2004 September and 2005 July 
(bottom panel of Fig.\,\ref{FUIST}) also show compact features to the NW,  
but the insignificance of the proper motion from these three observations, 
$17\pm28$ mas~y$^{-1}$, shows that `NW' is {\em not} a dust feature moving 
from the star or a re-appearance of MTD's `Feature B', which had a similar P.A. 
The [3.99] and $M^{\prime}$ 2004 June images also show a concentration at the 
same P.A. as `C1' but closer to the star, but this is not seen in the later images. 
The radial distances of all these concentrations show a strong correlation with 
the wavelength of the observation, and the average distance is proportional to 
wavelength, being about 15 per cent greater than the radius of the first Airy 
ring ($1\farcs635\lambda$/D), so we consider them to be instrumental artefacts. 
This is a characteristic bias of MEM-deconvolved images of a point source 
embedded in extended emission (Monnier 2003) and are seen only in our later 
images because of the relative faintness of the dust emission features.

The positions of `C1' and `C0' are included in Table \ref{Kpos}, and confirm 
the continued motion of the southern dust feature(s) (Fig.\,\ref{Fmotion}). 
The `arm' is seen in the 3.99-$\umu$m images but is getting lost in the noise, 
and it is not possible to identify `E' with enough confidence to track its 
position. 
The 2004 September and 2005 July images follow the same pattern: confirmation 
of the expansion of the southern dust-emission peaks (Table \ref{Kpos}, 
Fig.\,\ref{Fmotion}) and presence of the eastern `arm', but no concentration 
corresponding to `E'. 

We examined the UIST images for traces of the dust formed in the 1993 episode, 
such as those seen in the 12.5-$\umu$m image. Only the $M^{\prime}$ image 
observed on 2004 September 18, when the water vapour content above the telescope 
was very low, shows emission near the expected position (Fig.\,\ref{FGem}, 
a direct image without MEM-reconstruction to avoid possible introduction of 
artefacts). 
We observed a faint patch at P.A. $216\degr$, close to that of `C1', located 3568 
mas from the star. If this is dust formed in the 1993 episode, the mean proper 
motion over the intervening 1.46~P would be 308 mas~y$^{-1}$, smaller than the 
319 mas~y$^{-1}$ for the corresponding feature in the 12.5-$\umu$m image observed 
in 2003 and the average proper motion of `C1', $326\pm7$ mas~y$^{-1}$) from our 
2002--5 observations (Table \ref{Tmotion}), suggesting that the dust may be slowing 
down. The outskirts of the central image shows extended emission to the SW, which 
we have marked `F' at P.A. = $155\degr$, at a distance of 2694 mas. If this is dust 
formed in the 1993 episode, the mean proper motion would be 232 mas~y$^{-1}$, 
close to that of corresponding feature in the 12.5-$\umu$m image and mean 
proper motion of `C0' ($244\pm9$ mas~y$^{-1}$, Table \ref{Tmotion}), so we believe 
this is the remnant of `C0' formed in the 1993 dust formation episode.

\subsection{Proper motions of the dust emission features} 
\label{SMotion} 

\begin{figure}                                   
\includegraphics[clip,width=8cm]{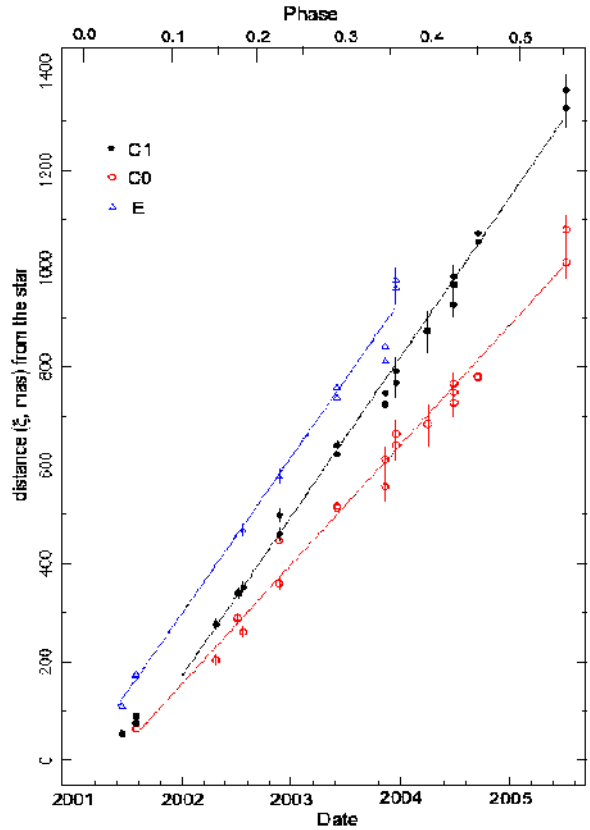}
\caption{Angular distances ($\xi$) of prominent dust emission features `C1', 
`C0' and `E' plotted against date (Table \ref{Kpos}). The straight lines 
are linear fits to our data and the Keck observations of `Feature E', 
which we identify with our `E'.}
\label{Fmotion}
\end{figure}

\begin{table}                                 
\caption{Proper motions, projected velocities and `start' ($\xi=0$) dates 
and phases of persistent dust emission features.}
\begin{tabular}{lllll}
\hline
Knot  & P. M.       & Proj. vel.   &  Date ($\xi=0$)  &  Phase ($\xi=0$)\\
      & (mas/y)     & (km/s)       &                  &                 \\
\hline
`C1'  & 326$\pm$7   & 2860$\pm$60  & 2001.50$\pm$0.08 & 0.050$\pm$0.011 \\
`C0'  & 244$\pm$9   & 2140$\pm$76  & 2001.37$\pm$0.14 & 0.034$\pm$0.018 \\
`E'   & 316$\pm$12  & 2767$\pm$105 & 2001.06$\pm$0.12 & 0.995$\pm$0.015 \\
\hline
\end{tabular}
\label{Tmotion}
\end{table}  

The positions of the labelled dust concentrations are used to characterize 
the motion of the dust clouds away from the star. 
As can be seen from Table \ref{Kpos}, the P.A.s of `C1' measured from our 
images fall in a narrow range, having an average $216\degr\pm2\degr$ from 
19 observations over three years. The P.A.s of `C0' and `E' have averages 
$154\degr\pm6\degr$ and $112\degr\pm4\degr$ respectively. The scatter in 
P.A.s is comparable to the observational uncertainties, so we believe that 
we are seeing the same dust concentrations in radial expansion, greatly 
extending the demonstration of homologous expansion of the dust from 
the early images by MTD. This is shown in Fig.\,\ref{Fmotion}, where the 
radial distances of `C1', `C0' and `E' from the central star are plotted. 

The average of the P.A.s measured for `E' is very close to those of 
`Feature E' given by MTD and measured by ourselves from their published 
3.08-$\umu$m image, so we believe that these are the same physical entity  
and included both in the fit to determine the proper motion. 

We also identify a dust feature in the Keck 2.21-$\umu$m image with our 
concentration `C0'. Its P.A. matches our mean P.A. for `C0' and its radial 
distance fits the linear expansion (Fig.\,\ref{Fmotion}) found for `C0'. 
We do not see it in the  Keck 2.26- and 3.08-$\umu$m images observed on 
the same date (Tuthill et al. 2003), although the 2.26-$\umu$m image shows 
an E--W structure consistent with the `bar'. 

We did not plot $\xi$ for MTD's `Feature D' and our `D' from the INGRID image 
(Fig.\,\ref{FPharo}), and are not confident that they are the same entity 
since the dispersion in P.A. is much greater than those of the persistent 
concentrations. Also, we did not recover `D' in any of our subsequent 
observations, nor did we observe concentrations corresponding to MTD's 
Features `A' or `B'. 

Linear fits to the distances give the proper motions in Table \ref{Tmotion}.
Those of `E' and `C1' are very similar, while that of `C0' is significantly 
less. This could be caused by a lower physical velocity or a different 
inclination angle of its trajectory, which can in principle be tested by 
comparison of its fading rate with that of `C1' (Section \ref{SPimages}).

We transform the observed proper motions to the transverse velocities given 
in Table \ref{Tmotion} assuming a distance of 1850~pc. The proper motions of 
`C1' and `E' are close to the terminal velocity of the WC7 stellar wind 
(2860 km~s$^{-1}$, Eenens \& Williams 1994),  
but this is fortuitous: we expect the velocities of the dust grains to 
be the sums of their terminal drift velocities and those inherited from 
the wind in which they formed. The relative constancy of the proper motions 
points to constant terminal drift velocities in a $\rho(r)\propto r^{-2}$ 
environment where the acceleration and drag forces balance:

\[
\rho(r)u^{2} = \frac{\sigma}{c}
\left(\overline{Q}_{\rmn{pr}}(a,T_{\rmn{O}})T_{\rmn{O}}^4 R_{\rmn{O}}^2
+ \overline{Q}_{\rmn{pr}}(a,T_{\rmn{W}})T_{\rmn{W}}^4 R_{\rmn{W}}^2\right)
r^{-2}. 
\]

There may a higher density region about 800 mas from the stars in the direction 
of `E', which slowed it down in 2003 as indicated by its radial distance 
measured from both the images observed in 2003.86. Otherwise, the constancy of 
the proper motions implies that the circumstellar environment through which the 
dust is moving does have a $\rho \propto r^{-2}$ density distribution, as 
produced by a constant velocity wind, with any irregularities in the 
interstellar medium long since swept up into the ring observed with {\em IRAS}.

The initial velocity of the dust must be that of the dense wind in which it 
condensed. The compressed wind material in a wind-collision region (WCR) flows 
slowly near the stagnation point and accelerates as it moves further from the 
stars. Cant\'o, Raga \& Wilkin (1996) derived formulae for the velocity of the 
shock-compressed wind flowing in a thin shell along the contact discontinuity 
in terms of the wind velocities of the two stars and the wind-momentum ratio, 
$\eta = (\dot{M} v_{\infty})_{\rm O} / (\dot{M} v_{\infty})_{\rm WR}$. 
The asymptotic value for the velocity of the compressed WC7 wind in WR\,140, 
using the stellar winds of the WC7 and O5 stars from Paper~1 and, e.g., 
$\eta$ = 0.1 (VWA) would be 2373 km~s$^{-1}$, about 80 per cent of the WC7 
wind terminal velocity.  

It is uncertain how well the `thin shell' analytic model can describe real, 
turbulent wind collisions as discussed by, e.g., Walder \& Folini 2002, but 
there is observational evidence for sub-terminal velocities of compressed WC 
star wind material in WCRs. Emission from such material is observed in 
`sub-peaks' on the broad emission lines of low-excitation ions, notably the 
He\,{\sc i} $\lambda$1.083$\umu$m and C\,{\sc iii} $\lambda$5696\AA\ lines. 
The evolution of the profile of the latter line in WR colliding wind systems 
has been studied by several authors, e.g. L\"uhrs (1997), Hill, Moffat \&
St-Louis (2002) and, in the case of WR\,140, MM03. 
The changing profile of the line as the orbits progress has been modelled 
geometrically in terms of emission by material flowing along 
the surface of a cone (approximating the WCR far from the stars) at constant 
velocity, giving values for the cone opening angle and flow velocity. The 
flow velocities derived are typically $\sim$ 75 per cent of the wind terminal 
velocities, with that for WR\,140 being $2300\pm500$ km~s$^{-1}$ (MM03). 
We therefore suggest that the dust grains formed in WR\,140 had initial 
velocities $\sim$ 2400 km~s$^{-1}$ and were accelerated by radiation pressure 
to achieve their observed constant velocities. This implies drift velocities 
as high as $\sim$ 500 km~s$^{-1}$ through the local compressed wind, which 
are rather high for survival of the grains against sputtering by He and C 
ions.

\subsection{`Start dates' and dust-formation time-scales}  
\label{SStart} 

The abscissae (Table \ref{Tmotion}) are related to the phases at which the 
dust-forming plasma was initially compressed in the densest part of the WCR, 
near the stagnation point between the stars. The stellar winds are compressed 
in the WCR throughout the orbit, so we use the term `dust-forming plasma' to 
refer to wind sufficiently compressed to nucleate dust as it flows further away 
from the stars. We refer to the dates in (Table \ref{Tmotion}) as `start dates' 
for convenience, but this would be true only if the concentrations had constant 
velocities between their formation and our first measurements. We expect the 
dust-forming plasma to have accelerated within the WCR and the newly-formed dust 
to have accelerated further by radiation pressure, so the dates when the 
dust-forming plasma started its motion are earlier than those in 
Table \ref{Tmotion} from extrapolation to $\xi=0$. 
This is evident from the light curves, which showed that the nucleation of new 
dust was complete by $\phi = 0.03$ (Section \ref{SPhot}, Fig.\,\ref{Kmax}), 
before the extrapolated $\xi=0$ phases of `C0' and `C1', the strongest dust 
features which would have dominated the infrared photometry. 
However, the interval between the first (`E') and last (`C1') start dates 
($0.055\pm0.018$~P) is about twice the duration (0.03~P) of dust nucleation 
inferred from the light curves. 
Although the intervals can be reconciled within the errors, the difference 
suggests that the delays between compression of the dust-forming plasma and 
dust nucleation are different for the different features. The alternative is 
that the timing of the 2001 eruption was different from the 1985 and 1993 events, 
on which the light curves are based. We do not have an infrared light curve for 
the 2001 eruption, but the timing is loosely constrained by the 2000.69 mid-IR 
photometry, showing that dust-formation had not started at $\phi = 0.95$, and 
the 2001.25 near-IR photometry showing dust emission at $\phi = 0.02$. 
Also, we know from the dilution of the emission lines in the IR spectra 
observed by VWA that dust formation began some time 
between 2001 January~3 and March~26 ($\phi = 0.989-0.017$), and that between 
April 28 and May 21 ($\phi = 0.028-0.036$) the near-IR emission was fading, 
indicating that nucleation had stopped. This is consistent with the previous 
eruptions. 
Also, the longer wavelength images (Fig.\,\ref{FGem}, Section \ref{SImages}) showing 
dust features from the 1993 eruption located at the same P.A.s as the strongest 
features `C0' and `C1' from the 2001 eruption, shows no reason to believe that the 
timing of the 2001 dust formation differered from those of the previous episodes. 

As the nucleation radius ($\sim$ 125 au) is large compared with the dimensions 
of the orbit (c. 16 au), the flow times will not vary significantly with the 
movement of the WCR with phase. 
The implication of the difference in delays is that nucleation does not always 
occur as soon as the dust-forming plasma reaches the nucleation radius. 
If it is the case that the nucleation of the earliest dust occurred 
less promptly than that of `C1' or `C0', thereby shortening the nucleation 
interval, the acceleration of the older dust by radiation pressure would also 
have been delayed, and its `start date' would have been even earlier. 
This is considered further when we compare the distribution of dust around the 
star with the projected motion of the WCR around the orbit.

\subsection{Photometry from the images}  
\label{SPimages} 

Contemporaneous imaging at different wavelengths allows estimation of the IR 
colours of the circumstellar features relative to those of the star. We measured 
the fluxes using software apertures on the images. The results from such photometry 
must be viewed with caution because, as Monnier (2003) points out, photometry of 
MEM-reconstructed images is necessarily biased, leading to systematic lowering of 
the estimated fluxes of compact sources. We used a small aperture (typically 0.5 
arcsec) on the star excluding, as far as possible, the dust emission to calibrate 
the images and a larger aperture, including the stellar and dust emission (`System'), 
to simulate the conventional photometry for a consistency check. 
As the system evolved, we used successively larger apertures for the `System' 
and dust features but always took care to match the positions and numbers of 
pixels included in the corresponding aperture on each set of 
images observed at the same epoch to get the best estimate of the colours. 
In addition to `C0' and `C1', measured through 0.4--0.5-arcsec apertures, we 
included them and the emission between them in a `bar' aperture, increasing from 
0.8$\times$0.3 arcsec (E-W$\times$N-S) at phase 0.23 to 1.5$\times$0.4 arcsec at 
phase 0.56.  
Most images show emission between `C0' and `C1', sometimes concentrated in a knot, 
suggesting that the `bar' is a single structure having 2--3 regions of higher 
(projected) density, perhaps emphasised in the images by the reconstruction process. 
This is supported by the close similarity of the `start' dates of `C0' and `C1', and 
not ruled out by the difference in their proper motions. We also used rectangular 
apertures to measure flux from the `arm', increasing from 0.3$\times$1.0 arcsec 
(E-W$\times$N-S) at phase 0.23 to 0.4$\times$1.7 arcsec at phase 0.46. These 
included `E', which appears from its earlier `start' date to be a separate 
structure formed earlier. This is consistent with the Keck images, which show 
a tight concentration South of the star and a more diffuse one to the East, 
about 120 mas more distant from the stars.

The results from the UIST images are given in Table \ref{photU}, showing that all 
the features are significantly redder than the star, as expected from heated 
circumstellar dust. In the seven months between the $\phi$ = 0.23 and 0.29 
observations, the emission from the features fades relative to the star and, 
in most cases, becomes redder, as expected from cooling dust. 
The ($nbL^{\prime}$--[3.99]) colours in the 2003 June image appear to be 
anomalously red, being comparable to those from the 2004 June image observed a 
year later, and much greater than those from the 2002 November image, 
probably due to problem with the reconstruction of the 2003 $nbL^{\prime}$ image.

\begin{table}
\caption{Photometry from the UIST images through software apertures centred on the 
star and various dust emission features. 
The `System' aperture was centred on the star and aimed to include stellar 
and emission feature fluxes. The apertures for the `Bar' and `Arm' were 
rectangular and their dimensions are given in the text. The $nbL^{\prime}$, [3.99] 
and $M^{\prime}$ magnitudes and the colours are given relative to the central star.}
\begin{tabular}{lcccccc}
\hline
Feature      & Phase  & $nbL^{\prime}$ & [3.99] & $M^{\prime}$ & ($nbL^{\prime}$ & ([3.99] \\
             &        &                &        &              & --[3.99])  &  --$M^{\prime}$) \\
\hline               
Star         &        & 0.0            & 0.0    & 0.0          & 0.0          & 0.0   \\
`C1'         & 0.23   & 1.0            & 0.5    &              & 0.5          &       \\
`C1'         & 0.29   & 2.0            & 1.3    &              & 0.7          &       \\
`C1'         & 0.43   & 3.7            & 2.9    & 2.8          & 0.8          & 0.1   \\
`C1'         & 0.46   &                & 3.1    & 2.9          &              & 0.2   \\
`C1'         & 0.56   &                & 3.9    & 3.6          &              & 0.3   \\
             &        &                &        &              &              &       \\
`C0'         & 0.23   & 0.7            & 0.3    &              & 0.4          &       \\
`C0'         & 0.29   & 2.2            & 1.3    &              & 0.9          &       \\
`C0'         & 0.43   & 3.4            & 2.8    & 2.5          & 0.6          & 0.3   \\
`C0'         & 0.46   &                & 3.2    & 2.8          &              & 0.4   \\
`C0'         & 0.56   &                & 3.9    & 3.5          &              & 0.4   \\
             &        &                &        &              &              &       \\
`Bar'        & 0.23   & -0.1           & -0.5   &              & 0.4          &       \\
`Bar'        & 0.29   & 1.2            & 0.4    &              & 0.8          &       \\
`Bar'        & 0.43   & 2.3            & 1.7    & 1.6          & 0.6          & 0.1   \\
`Bar'        & 0.46   &                & 1.9    & 1.6          &              & 0.3   \\ 
`Bar'        & 0.56   &                & 3.0    & 2.3          &              & 0.7   \\ 
             &        &                &        &              &              &      \\
`Arm'        & 0.23   & 2.0            & 1.6    &              & 0.4          &      \\
`Arm'        & 0.29   & 2.6            & 1.9    &              & 0.7          &      \\
`Arm'        & 0.43   & 5.2            & 3.4    & 3.9          & 1.8          & -0.5 \\
`Arm'        & 0.46   &                & 3.2    & 2.8          &              & 0.4  \\ 
             &        &                &        &              &              &      \\
System       & 0.23   & -1.0           & -1.2   &              & 0.2          &      \\
System       & 0.29   & -0.4           & -0.7   &              & 0.3          &      \\
System       & 0.43   & -0.2           & -0.3   & -0.3         & 0.1          & 0.0  \\
System       & 0.46   &                & -0.4   & -0.4         &              & 0.0   \\
System       & 0.56   &                & -0.1   & -0.2         &              & 0.1   \\
\hline
\end{tabular}
\label{photU}
\end{table}

\begin{figure}                                     
\includegraphics[clip,width=8cm]{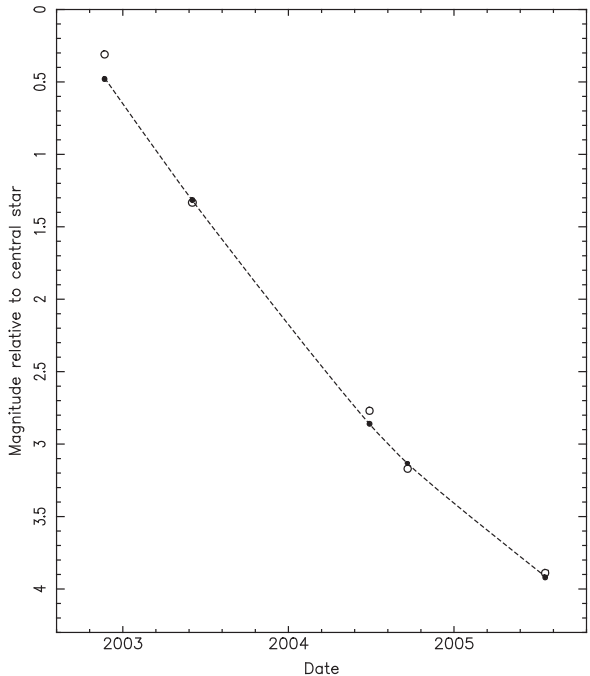}
\caption{Comparison of [3.99] magnitidues of `C1' ($\bullet$ and broken line) 
and `C0' ($\circ$) relative to the central star.}
\label{FCC0399}
\end{figure}

We observed the fluxes from `C0' and `C1' to test from their relative cooling rates 
whether their significantly different proper motions are caused by differences in 
the inclination angles of their trajectories or real differences in expansion 
velocity. If the 75 per cent lower proper motion of `C0' is attributable to a lower 
velocity, `C0' would always be closer to the stars. The dust temperature is expected 
to fall off as $T_{\rmn{g}} \propto r^{-0.38}$ (Section \ref{SPhot}) so the dust in 
`C0' should be about 12 per cent hotter than that in `C1'.  This leads to an emissivity 
ratio varying from 1.8 to 2.2 at 3.99$\umu$m (the wavelength at which most observations 
are available) as the dust cools so that, whatever the ratio of dust masses in `C0' 
and `C1', the [3.99] magnitude difference between the concentrations should widen 
as the dust fades. The fading of the [3.99] magnitudes of `C1' and `C0' is compared 
in Fig.\,\ref{FCC0399} and there is no evidence that `C1' is fading more quickly. 

Modelling the ($nbL^{\prime}$--[3.99]), ([3.99]--$M^{\prime}$) and ([7.9]--[11.5]) 
colours for the relevant temperatures leads us to expect that `C0' should be 
$\simeq$ 0.1 mag bluer than `C1' if it were moving more slowly, but the observed 
colours (Table~\ref{photU}--\ref{photG}) do not support this: 
($nbL^{\prime}$--[3.99]) are bluer, ([3.99]--$M^{\prime}$) are redder, and 
([7.9]--[11.5]) are the same. We therefore consider it more likely that `C0' and `C1' 
are at the same temperature and distance from the star, and that the difference in 
proper motion is attributable to a difference in the inclination angles of their 
trajectories of to our line of sight. 
This does not rule out the possibility that they might be concentrations in the 
same physical structure.

The dust photometry can be calibrated via the star. From the pre-outburst 
photometry (Section \ref{SPhot}), we used $M^{\prime}=4.53$ and [12.5] = 3.25, 
and interpolated $nbL^{\prime}=4.85$, [3.99] = 4.75, and [7.9] = 3.95 for 
the dust-free star. 
The System (star+dust) magnitudes provide a consistency check. 
For example, the differences in Table \ref{photU} give 
System magnitudes $nbL^{\prime}$ = 3.9 and [3.99] = 3.6 in 2002 November 
($\phi$ = 0.23) and $nbL^{\prime}$ = 4.4 and [3.99] = 4.0 in 2003 June 
($\phi$ = 0.29), which compare well with the $L^{\prime} \simeq 3.93$ and 
$4.15$ for these two phases on the photometric light curve. 
Similarly, the differences in Table \ref{photG} give System magnitudes 
[7.9] = 2.7 and [12.5] = 2.2 at phases 0.33 and 0.34, consistent 
with the photometry at phase 0.34:  [8.75] = 2.8 and [12.5] = 2.3.
As noted above, the calibrated 2004--5 UIST images gave $M^{\prime}$ magnitudes 
of the combined system in excellent agreement with the $M/nbM$ light curve. 
They agree also with the system magnitudes derived from the pre-outburst 
calibration and the differences in Table \ref{photU}.

The imaging observations cover too short a baseline in wavelength for 
determining the temperature of the dust, but we can check for consistency 
using the average dust temperature derived from fitting the 2--20 $\umu$m 
SEDs. At phase 0.26 (between the 2002 and 2003 imaging observations), the  
fit gave $\langle T_{\rmn{g}} \rangle \simeq 700~K$ and emission from dust 
at this temperature has ($nbL^{\prime}$--[3.99]) = 0.35, which is 0.25 mag 
redder than the star. The ($nbL^{\prime}$--[3.99]) differences from the 2002 
November ($\phi = 0.23$) image are slightly greater than this, but those 
from the 2003 June ($\phi = 0.29$) image significantly greater, as noted above. 
The final UIST image was also observed at a phase ($\phi = 0.56$) close to 
one at which we have photometry out to 20~$\umu$m. 
That SED gives $\langle T_{\rmn{g}} \rangle \simeq 530~K$, which has 
([3.99]--$M^{\prime}$) = 0.6, reasonably consistent with the differences in 
Table \ref{photU}.

\begin{table}
\caption{Photometry of features and the star+dust system from the Gemini 
images. The [7.9], [12.5] and [18.5] magnitudes and ([7.9]--[12.5]) 
colour are relative to the central star.}
\begin{tabular}{lccccc}
\hline
Feature & Phase &  [7.9]  &  [12.5] & [18.5] & ([7.9]--[12.5])\\
\hline         
Star    &       & 0.0     & 0.0     & 0.0    & 0.0    \\
`C1'    & 0.35  & 0.8     & 1.1     &        & -0.3   \\
`C1'    & 0.36  & 0.9     & 1.1     & 1.3    & -0.2   \\
`C0'    & 0.35  & 0.4     & 0.7     &        & -0.3   \\
`C0'    & 0.36  & 0.6     & 0.8     & 1.2    & -0.2   \\
`Bar'   & 0.35  & -0.4    & -0.1    &        & -0.3   \\
`Bar'   & 0.36  & -0.4    & -0.1    & 0.2    & -0.3   \\
`Arm'   & 0.36  & 1.2     & 1.4     & 1.3    & -0.2   \\
System  & 0.36  & -1.2    & -1.0    & -0.9   & -0.2  \\
\hline
\end{tabular}
\label{photG}
\end{table}

Photometry through software apertures on the Michelle images is given in Table 
\ref{photG}. 
Counter-intuitively, the dust is {\em bluer} in ([7.9]--[12.5]) than the star. 
This is correct. It occurs because, at the time of the observations, the dust 
emission peaked at a wavelength shorter than those of the observations, and the 
Rayleigh-Jeans tail of the Planckian dust spectrum is steeper than the spectrum 
of the free-free emission from the stellar wind. From fitting the 3.5- to 
20-$\umu$m SED for phase 0.34, we get $\langle T_{\rmn{g}} \rangle \simeq 600~K$, 
giving a model colour ([7.9]--[12.5]) $\simeq$ 0.25. From the pre-outburst 
photometry (Section \ref{SPhot}), we interpolate ([7.9]--[12.5]) = 0.75 
for the star, which is about 0.5 mag redder than the dust features, comparable 
with our images (Table \ref{photG}). The calibrated System colours from the images 
are ([7.9]--[12.5]) = 0.5--0.6, consistent with the ([8.75]--[12.5]) = 0.5 
observed at phase 0.33.

We examined the evolution of the dust mass in the most persistent feature, the 
`bar', and used the [3.99] magnitudes in Tables \ref{photU}, the longest dataset 
and at a wavelength near the dust SED peak. We calculated 
dust emissivity from grain temperatures T$_{\rmn{g}}$ estimated by interpolating 
amongst the $\langle T_{\rmn{g}} \rangle$ determined from fitting the multi-band 
$H$ to [19.5] data at those phases at which photometry over the full wavelength 
range was available. The `bar' dominates the dust emission, so this should give 
reasonable estimates of its temperature. 
The  relative dust masses derived (Table \ref{dmass}) are independent of grain 
emission coefficient but do rely on the star being constant at 3.99 $\umu$m. 
They show that the mass of dust falling steadily during the period covered, 
but at about twice the rate at which the total dust mass was found to fall 
in the same phase interval from modelling the photometry (cf. Section \ref{SPhot}) 
to which the `bar' flux makes a significant contribution. This suggests that, as 
the dust emission fades to approach the noise level in the images, we are 
progressively losing flux in the reconstructed images and that the reduction of 
dust mass in Table \ref{dmass} is an overestimate.

\begin{table}
\caption{Evolution of the relative dust mass, M$_{\rmn{d}}$, in the `bar', 
along with the measured [3.99] difference and adopted temperature.}
\begin{tabular}{llccc}
\hline
Date    & Phase & [3.99] & T$_{\rmn{g}}$ & Rel. M$_{\rmn{d}}$ \\
\hline
2002.89 &  0.23 & -0.50  &  730          & 1.00              \\
2003.42 &  0.29 &  0.44  &  665          & 0.68              \\
2004.49 &  0.43 &  1.71  &  565          & 0.55              \\
2004.71 &  0.46 &  1.87  &  555          & 0.54              \\
2005.52 &  0.56 &  2.98  &  530          & 0.26              \\
\hline
\end{tabular}
\label{dmass}
\end{table}

The emission from the dust `arm' is fainter than that from the `bar' in all 
the images in which emission from both could be measured, pointing to a lower 
mass of dust in the `arm'. The emissivity will also be lower if the dust is 
cooler than that in the `bar', which will be the case if it is more distant 
from the stars, as is suggested by the projected distances of the features 
(Table \ref{Kpos}). This is supported by the ($nbL^{\prime}$--[3.99]), 
([3.99]--$M^{\prime}$) and ([7.9]--[12.5]) colours of the `arm', which are 
slightly redder or the same as those of the `bar'. The baselines are too 
short to determine grain temperatures from fitting the colours, so we use the 
radiative equilibrium temperature ratios, $T_{\rmn{g}} \propto r^{-0.38}$, 
assuming that the ratios of the distances of the `arm' and `bar' from the 
stars was equal to that of the projected distances (Table \ref{Kpos})  
measured from the images, to estimate temperatures. We use these to form 
synthetic colours and check them for consistency with those observed.  

The projected distances of `E' and `C1' at the times of the 2002 and 2003 UIST 
observations suggest that the dust in the `arm' would be about 50~K cooler than 
that in the `bar'. This would account for only 0.4 mag of the difference in 
[3.99] and 0.04 mag in ($nbL^{\prime}$--[3.99]). The colour difference is 
consistent with those in Table \ref{photU}. The balance of the flux differences 
from the 2002 and 2003 observations suggests that mass of dust in the `arm' was 
lower than that in the `bar' by a factor of $\simeq$ 4.  Similarly, we get a 
40-K difference in `bar' and `arm' temperatures at the time of the 2003 Gemini 
observations, accounting for 0.23 mag of the difference in [7.9] and 0.07 
mag in ([7.9]--[12.5]). 
The measured differences in ([7.9]--[12.5]) are greater (0.1 and 0.2 mag, 
Table \ref{photG}, implying a greater temperature difference, for which we 
adopt 60~K. This accounts for 0.35 mag (at 7.9 $\umu$m) of the differences 
between `bar' and `arm' fluxes. The balance of the measured flux differences 
indicates a lower dust mass in the `arm', by a factor of $\simeq$ 3.
Finally, a 50-K difference in `bar' and `arm' temperatures at the time of the 
2004 September UIST observation would account for 0.7 mag of the difference 
in [3.99] and 0.09 mag in ([3.99]--$M^{\prime}$). The latter is consistent with 
the observed colours (Table \ref{photU} and the balances of the measured 
[3.99] and $M^{\prime}$ differences give a dust mass ratio of $\simeq$ 4.
Owing to the possible loss of flux in the fainter reconstructed images, 
this ratio is probably overestimated; but all the observations, going back to 
the 2002 images, point to fainter dust emission in the `arm' so there is 
certainly some mass difference.

The distances of the dust features from the 1993 dust-formation episode, `F' 
and `G' observed with Gemini at 12.5$\mu$m (Fig.\,\ref{FGem}) are about 
$4.1\times$ those of `C0' and `C1', giving radiative equilibrium temperatures 
$\langle T_{\rmn{g}} \rangle \simeq 350$~K for `F' and `G'. At 12.5 $\umu$m, such 
dust has an emissivity about 19 per cent that of the 600-K dust from the 2001 
episode. If the mass of dust in `G' was the same as that in the corresponding 
feature, `C1', it would be 1.8 mag fainter. The observed magnitude difference is 
nearer 3.4 mag, suggesting that much of the dust from the 1993 episode has been 
lost, consistent with the fall in the mass of dust in the `bar', but this 
effect may be overestimated as noted above given the faintness of the emission.
Similarly, the corresponding feature in the 2004 September UIST $M^{\prime}$ image 
would have $\langle T_{\rmn{g}} \rangle \simeq 340$~K and an emissivity at this 
wavelength of only 1 per cent that of the 550-K dust from the 2001 epsiode. 
Comparison with the measured magnitude difference suggests that half the original 
dust had been lost but this is very uncertain. The emissivity ratio at shorter 
wavelengths is even less, only 0.5 per cent at 3.99 $\umu$m, accounting for our 
not observing the `1993' dust in the [3.99] images.

\section{Relation of the dust images to the orbit and CWB paradigm}  
\label{SRelCWB}  

\subsection{Changing configuration of WR\,140 during dust formation}  
\label{SConfig} 

\begin{figure}                                   
\includegraphics[clip,width=8cm]{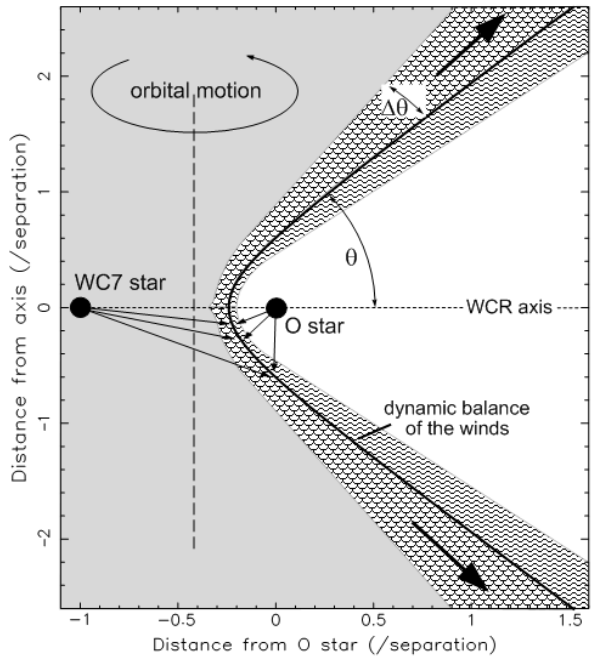}
\caption{Sketch of the wind-collision region (WCR) relative to the stars; 
the observer is in the plane of the orbit, perpendicular to the WCR axis. 
The winds of the WC7 and O stars are separated by a contact discontinuity 
approximated at large distances by a cone whose opening angle, $\theta$, 
depends on the ratio of the momenta of the two winds. On either side of it 
are surfaces where the WC7 and O stellar winds are shocked. This effect 
diminishes in the outer regions of the WCR as the angles of incidence of 
the undisturbed WC7 and O5 winds on the WCR get smaller.
Note that our definition of $\Delta\theta$ differs from that of 
Eichler \& Usov (1993), which includes the shocked regions on both sides of 
the contact discontinuity.}
\label{Config}
\end{figure}

The CWB paradigm has dust forming from the compressed wind in the WCR and 
expanding radially from the stars. The WCR at large distances from the stars 
may be approximated by a hollow cone of opening angle $\theta$ and thickness 
$\Delta\theta$ (cf. Fig.\,\ref{Config}), following analytical models of, 
e.g., Eichler \& Usov (1993) and Cant\'o et al. The WCR is assumed to be 
symmetric about its axis, the line of centres of the stars. The dust extends 
($\theta$+$\Delta\theta$) above and below the orbit, and this quantity could 
be determined directly from observations if the orbit was highly inclined. 
The WCR moves with the stars in their orbital motion, spreading the dust 
round the stars, and would produce a `pinwheel' if the dust formation was 
continuous. In the case of WR\,140, which makes dust for only a small fraction 
of its period, the distribution of dust in the plane of its orbit is determined 
by ($\theta$+$\Delta\theta$) and the angular movement of the WCR during 
the phase interval when the WCR was producing dust-forming plasma (i.e. suitably 
compressed wind), and we expect to observe only an arc of a dust pinwheel. 
After determining the orientation of the orbit on the sky, Dougherty et al.\, 
(2005) noted that the O5 star was NW of the WC7 star at the time of periastron, 
and commented on the paucity of dust in that direction, where they expected 
dust to form. Indeed, our images show the dust to be located in 
a number of clouds spread most of the way around around the stars, but not 
to the NW, where it was seen only in the early Keck images of MTD.  

Owing to the high orbital eccentricity (0.881, MM03), the position angle of 
the WCR changes very rapidly around periastron passage, so that the placement 
of the dust plume around the system is particularly sensitive to the timing 
of the production of the dust-forming plasma. 
For example, the P.A. of the O5 star relative to the WC star moves through more 
than one-quarter of its orbit, from $356\degr$ to $246\degr$, in only 0.01P 
($\phi = 0.995-0.005$), and through three-quarters of its orbit in only 0.04P. 
These ranges are very sensitive to the orbital eccentricity itself, because it 
is so high. They account for the spreading of the dust around much of the orbit 
despite the short duration of dust formation. 

Another consequence of the highly variable orbital velocity around periastron 
is that the density along the plume varies sharply with P.A. Even if dust forms 
at a constant rate, the density on the sky of dust condensed from dust-forming 
plasma originating at periastron is more than ten times lower than that 
originating at phase 0.02, where the WCR sweeps round more slowly, and the dust 
is spread less thinly.

\subsection{Modelling the dust distribution}  
\label{SModel}  

As a baseline, we model the distribution of dust made by WR\,140 according to 
the CWB paradigm. For each phase between the start and end phases (or the phase 
of the model if that occurs sooner), we calculate the orientation of the WCR 
from the orbital elements of MM03 and Dougherty et al. The dust formed at that 
phase will form a ring whose location, size and orientation are defined by 
extending the WCR away from the stars in proportion to the interval since the 
start phase. This ring is then projected 
on a $200 \times 200$ grid on the sky to build up a 2-D map of the dust, which 
can be translated directly into emission as the dust is optically thin in the 
IR (Paper~1). For comparison with observations, the dust map is convolved 
with a Gaussian profile.

Besides the orbit, the principal parameters determining the dust distribution 
are the shape of the WCR and duration of the process. Other parameters are the 
efficiency with which dust nucleates from the plasma, the density distribution 
around the WCR, and the velocity of the dust.

The shape of the WCR is determined by the dynamical balance of the two stellar 
winds at the contact discontinuity. From the stagnation point on the axis, 
where the winds collide head-on, the WCR curves until, further from the stars, 
it can be approximated by a cone. If the stellar winds have reached their 
terminal velocities before collision, the value of the opening angle, $\theta$,  
depends only on the wind-momentum ratio,
$\eta = (\dot{M}v_\infty)_{\rmn{O5}}/(\dot{M}v_\infty)_{\rmn{WC}}$, 
and does not vary with stellar separation around the orbit. In this case, the 
shape of the WCR would remain the same while its proximity to the stars varied 
with the stellar separation. From consideration of the mass-loss rates of the 
WC7 and O5 stars based on X-ray and radio observations at $\phi = 0.837$, 
Pittard \& Dougherty (2006) derived a wind-momentum ratio $\eta = 0.02$, 
which gives $\theta=31\degr$. From spectroscopic observations near periastron, 
MM03 determined $\theta=40\pm15\degr$ from modelling the moving sub-peaks on 
C\,{\sc iii} and He\,{\sc i} emission lines, and VWA found 
$50\degr<\theta<60\degr$ (for $i=122\degr$) from the orbital variation of the 
absorption component of the $\lambda$1.083-$\umu$m He\,{\sc i} line.
We begin by adopting, as a round number, $\theta = 40\degr$, corresponding to 
$\eta = 0.046$.
 
Near periastron, the distance from the O5 star to the stagnation point 
is given by
\[
r_{\rmn{O5}} = \frac{\sqrt{\eta}} {\sqrt{1 + \eta}} D, 
\]
\noindent 
where $D$ is the stellar separation (Eichler \& Usov), and could become too small 
for the O5 stellar wind to have accelerated to its terminal velocity before 
collision. At periastron, $D = 0.12a$ and, if $\eta = 0.046$ and 
$R_{\rmn{O5}} = 26 R_{\odot}$ (as above), we have $r_{\rmn{O5}} = 2.9 R_{\rmn{O5}}$. 
For a simple $\beta$ wind velocity-law of the form
\[
v(r) = v_{\infty} \left( 1 - \frac{R_{\rmn{O5}}}{r} \right) ^ {\beta},
\]
\noindent
the O5 wind velocity would be only $0.66~v_{\infty}$ (assuming $\beta = 1.0$) when 
it reached the WCR at $r = r_{\rmn{O5}}$, and the WCR would contract around the O5 
star, with a smaller opening angle, $\theta$. The effect would be greater for a 
smaller wind-momentum ratio, e.g. $v(r_{\rm O5}) = 0.5~v_{\infty}$ for $\eta = 0.02$. 
Countering this effect is the possible radiative braking of the WC7 wind by 
the O5 star's radiation field, normally insignificant in WR\,140 (Gayley, 
Owocki \& Cranmer 1997), which might occur as the WCR moves closer to the O5 
star, so that a wind-momentum balance could still exist. Detailed modelling of 
the WCR under these conditions is beyond the scope of the present study; for 
which we recall the persistence of the C\,{\sc iii} sub-peak emission right 
through periastron (MM03), and of dust nucleation as suggested by the 
$K$-band photometry (Fig\,\ref{Kmax}), which indicate that the WCR does not 
collapse on to the O5 star. Given that the effects of incomplete acceleration 
of the O5 wind and radiative braking of the WC7 wind work in opposite 
directions, we assume $\theta$ to be constant during dust formation. 
Consequently, the fraction of the WC7 stellar wind going into the WCR and 
becoming available for dust formation would be independent of phase. 
We also assumed that the fraction of compressed wind that condensed dust 
was constant while this phenomenon occurred, i.e. that dust nucleation was 
a threshold phenomenon. 

The appropriate value of $\Delta\theta$ is unknown: Pittard \& Dougherty 
(2006) found values of $\Delta\theta \simeq 20\degr$ for a range of possible 
values of $\theta$, but this applies to adiabatic post-shock winds. Where 
there are radiative losses, which is the case with WR\,140 near periastron 
(MM03, VWA), $\Delta\theta$ will be smaller (Eichler \& Usov 1993), and we 
arbitrarily adopt $\Delta\theta = 10\degr$ here for this phase range. In 
principle, it could be much thinner, but instabilities from cooling could 
again increase its geometrical thickness (Folini \& Walder 2002).

The WCR moves with the stars in their orbit, the axis lagging behind the  
line of centres because of the orbital motion. Owing to the high eccentricity, 
the aberration angle, $\delta = \arctan(v_{\rmn{orb}}/v_{\rmn{wind}})$,   
resulting from the Coriolis force varies significantly round the orbit with 
the varying transverse velocity of the WCR, $v_{\rmn{orb}}$, taken to be that 
of the WCR relative to the WC7 star. 
Owing to the long orbital period and fast stellar winds in WR\,140, $\delta$ 
is expected to be small and only reach $5\degr$ at periastron. This is  
smaller than the corresponding angle, $\delta\phi = 40\degr\pm20\degr$ 
(assumed constant) found from fitting the emission-line subpeaks (MM03), 
and we used the aberration angle from the transverse velocity in our modelling.

The distribution of compressed wind and dust around the axis of the WCR is unknown. 
Models of an adiabatic WCR in $\gamma$~Vel show a greater density enhancement in 
the trailing edge (in the orbital plane) of the WCR than in the leading edge 
and this was predicted to carry over to the radiative case (Folini \& Walder). 
Similarly, Lemaster, Stone \& Gardiner (2007) found higher densities on the 
trailing edges of the WCR in their study of the effects of the Coriolis force 
on CWB interactions but, as none of these models was directly applicable to 
the case of WR\,140, we began by assuming an axisymmetric density distribution 
and consider deviations below. 

The compressed wind accelerates in the WCR from the stagnation point to the 
cone-shaped region, where it reaches an asymptotic velocity about 80 per cent 
that of the stellar winds (Section \ref{SMotion} above). The wind is assumed to 
continue ballistically on this trajectory. During some interval, dust condenses 
in the down-stream flow and is quickly accelerated by radiation pressure to its 
terminal drift velocity.

\begin{figure}                                       
\includegraphics[clip,width=8.4cm, angle=270]{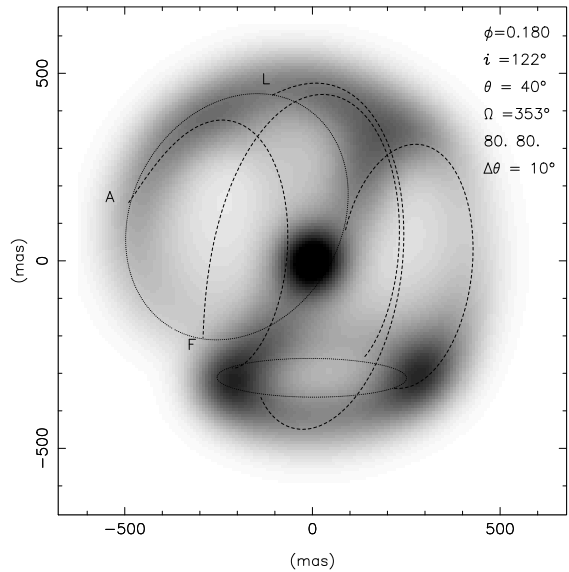}
\caption{Model dust image for $\phi$ = 0.18, near the epoch of the 2002 July 
observations (Fig.\,\ref{FPharo}), convolved with a Gaussian of 120 mas FWHM. 
The stellar image is attenuated to allow display of the dust emission. 
Superimposed is a frame comprising ellipses marking the projections of those 
regions of the WCR which the earliest (NE of the star) and latest (S of the star) 
dust to form would have reached by the model phase ($\phi$ = 0.18) and, 
joining them, arcs showing the loci of four points on these ellipses, that at 
the leading edge of the WCR in the orbital plane (marked `L'), the following 
edge (`F'), and maxima above (`A') and below (these are 
arbitrary) the plane. The dust is optically thin, so we see superposition of 
dust made at different phases and limb brightening of hollow structures.}
\label{Model88}
\end{figure}

\begin{figure}
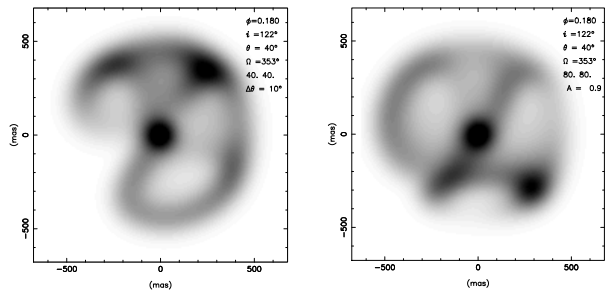
                                      
\includegraphics[clip,width=4.2cm, angle=270]{dm88n120.eps}
\hfill
\includegraphics[clip,width=4.2cm, angle=270]{masym9.eps}
\caption{Model dust images for $\phi$ = 0.18, showing the effects of 
(left) half the duration of dust formation ($\pm 40$~d, from the nucleation time)
and (right) an asymmetric dust distribution with $A=0.9$. 
Other parameters are the same as in Fig. \ref{Model88}.}
\label{ModelComp}
\end{figure}

As discussed in Section \ref{SStart}, there are two sets of data determining 
the orbital phases at which the WCR started and ceased to produce dust, each 
with its own assumptions and delays. On the one hand are the $\xi=0$ abscissae 
in Table \ref{Tmotion}, which are later than the true dates owing to the 
acceleration of the dust-forming plasma in the WCR, and then of the dust 
grains by radiation pressure. 
If these delays are the same for all dust features, the differences between 
the $\xi=0$ dates still hold, i.e. `E' and the `arm' were formed first, 
followed $0.039\pm0.023$~P and $0.055\pm0.019$~P later by `C0' and `C1'. 
The interval between the last two features is comparable to the uncertainties, 
and they may have been formed at the same time, which is compatible with 
the clockwise rotation of the system on the sky. 
On the other hand are the phases at which the near-IR light curves showed 
dust nucleation to be occurring 
($\phi \simeq 0.0-0.025$, Section \ref{SPhot}), giving half the duration. 

We begin by adopting the time-scale indicated by the proper motions. If the 
production of dust-forming plasma depends only on conditions such as local 
radiation field and pre-collision wind density, which depend on stellar 
separation, the $0.055$-P phase interval would occur symmetrically about 
periastron, i.e. running from $\phi=0.972$ to $\phi=0.028$ (80 d before and 
after periastron passage) when the local conditions would have been the same. 
We assume that, between these phases, the rate of compression of the 
dust-forming plasma was constant. We are aware that this runs counter to 
the observation that the mass of dust formed before periastron, the `arm',
is about one-quarter that formed afterwards but, in the absence of an 
obvious physical reason for this difference, continue with our assumption.

Using these parameters, we constructed a specimen model of the dust clouds 
for $\phi = 0.18$, near the time of the 2002 July observations. Comparison of 
the model (Fig.\,\ref{Model88}) and observed dust images (Fig.\,\ref{FPharo})
shows that the model reproduces the southern `bar', including the 
concentrations `C0' and `C1', reasonably well, but not the eastern `arm', 
and produces more dust to the north-west than seen in the observations. 
Owing to the high angular velocity of the WCR around periastron, the angular 
distribution of dust is very sensitive to the choice of interval during which 
dust-forming plasma is produced.  The placement of dust on the sky provides 
tight constraints on the orientation of the WCR when the dust-forming plasma 
was condensed and hence on the phases at which this process occurred. 
In particular, images using the duration of dust nucleation indicated by the 
light curves (40 d before and after periastron) are a very poor fit 
(cf. Fig.\,\ref{ModelComp}). 
Indeed, the fit can be improved slightly by increasing the production duration 
of dust-forming plasma by $\sim$ 0.01~P, or by adopting a higher orbital 
eccentricity, which increases the P.A. range covered in the same phase interval, 
but the discrepancy between model and observed dust to the NW of the stars 
requires re-examination of the input assumptions.

The model dust NW of the stars is not dust formed at periastron; dust from wind 
compressed around periastron would be spread with P.A. extending $\pm50\degr$
($\theta+\Delta\theta$) on either side of the projected P.A. at periastron. 
We tested this with a variant of the model which allowed for a brief pause in 
dust formation very close to periastron, not long enough to show up in the 
light curve (Fig\,\ref{Kmax}), as if the WCR was temporarily unable to produce 
dust owing to the proximity of the stars or weakening of the wind collision, 
but this did not remove the dust to the NW. With the aid of the overplotted 
wire frame in Fig.\,\ref{Model88}, we see that the model dust to the NW is 
attributable to the overlay of dust formed in the leading edge of the WCR 
before periastron and, to a lesser extent, the following edge after periastron. 
The observed paucity of dust in this direction suggests that we abandon the 
assumption that the dust is distributed uniformly around the axis of the WCR. 
This asymmetry could be related to the asymmetries found in WCR models referred 
to above, or it could be a consequence of asymmetry in the WC7 stellar wind such 
that the WCR moved through a region of rapidly falling wind density between 
$\phi \simeq 0.95$ and periastron, as if there were a high-density flattened 
region having P.A. $\simeq 100\degr$. A flattened wind for the WC7 star was 
proposed by White \& Becker (1995) to explain the variations in the radio 
emission, but this is not supported by polarimetric observations (MM03 and 
references therein). Also, the WCR would cross such a disk a second time 
$180\degr$ later in P.A. ($\simeq 280\degr$), leading to enhanced dust 
formation between P.A.s $240\degr$ and $320\degr$ -- which is not observed. 

We therefore ascribe the asymmetry to the WCR itself, and test this with the 
simplest parameterisation of non-uniform wind density around the WCR: a fraction 
$A$ of the dust is re-distributed around the axis of the WCR as $A\cos(\zeta)$, 
where $\zeta=0$ is the azimuthal angle corresponding to the trailing edge of the 
WCR in the orbital plane. A model image with this adjustment for $A = 0.9$ is 
presented in Fig\,\ref{ModelComp}. Although this does not reproduce the `arm', 
it represents an improvement over the uniform distribution model, and might be 
improved further with a physically more realistic azimuthal density distribution.

The model images (Figs \ref{Model88}--\ref{ModelComp}) show dust projected against 
the star. This is the earliest dust to form, made from the WCR when its axis was 
pointing towards us. This dust therefore lies in our line of sight to the stars, 
and is most probably that responsible for the optical eclipses observed by MM03 and 
previous observers shortly after periastron.

\section{Discussion} 
\label{SDiscuss} 

The unprecendented combination of high-resolution infrared imaging, multi-band 
infrared photometry and knowledge of its distance and the binary orbit in three 
dimensions provides a unique opportunity to examine the dust formation process 
by WR\,140 and assess some of the underlying assumptions, with implications 
for other, less well observed, dust-making Wolf-Rayet systems.

Consideration of the radiative equilibrium on the youngest dust and the   
projected distances of the dust features from the mid-2001 images suggests
that the optical properties of the grains formed by WR\,140 differ from 
those made by the WC9 stars, particularly the classical `pinwheel' WR\,104 
modelled by Harries et al. (2004), in having a lower ratio of UV-to-IR 
absorption coefficients. The physical conditions under which WR\,140 makes 
dust differ from those in WR\,104 -- e.g. the stronger UV radiation field, 
as both components of WR\,140 are hotter than the stars in in WR\,104 -- so 
such differences are unsurprising, and recall those in different samples of 
carbon grains formed in the laboratory (Andersen et al.). 
We also recognise from the comparison of their detailed observations of 
the WR\,104 dust plume by Tuthill et al. (2008) with the model developed by 
Harries et al. (2004) that the dust models based on radiative equilibrium in 
the stellar radiation field need refinement.

The images of the dust made by WR\,140 are very different from the classical 
`pinwheels' around the persistent dust-makers, more akin to `splashes', with 
patches of dust at similar distances and having a variety of position angles, 
moving away from the stars at constant rates. Extrapolation of the motions 
back to the stars suggests that the dust-forming plasma responsible for the 
features to the east was condensed about 0.055~P before that responsible for 
the features to the south, assuming that the acceleration was the same for all 
features. 
The spreading of the dust around most of the orbit can be reconciled with the 
formation of dust in a WCR rotating with the stars and the short duration of 
the production of the dust-forming plasma (0.055~P) by noting that the high 
eccentricity of the orbit causes the P.A. of the WCR to swing through 
almost two-thirds of its orbit ($233\degr$ from NE to S) in this interval. 
This is consistent with the dust to the E being formed before that to the S.

Calculations of model dust distributions on this basis have mixed success. 
They reproduce the most persistent dust feature, the southern `bar' with 
concentrations `C0' and `C1' at the two ends, but fail to the north and east. 
We have examined the effects on the images of changing model (e.g. $\theta$) 
and orbital parameters: most have little effect, apart from the orbital 
eccentricity and the phase range during which dust-forming plasma was produced. 
Owing to the rapid movement of the orientation of the WCR near periastron, the 
model images are very sensitive to the choice of the phases at which dust-forming 
plasma was condensed, and this provides a third, perhaps the tightest, constaint 
on the time-scale of dust formation. 
It is possible that current RV studies of WR\,140 will lead to small changes  
to the orbital parameters but these will have to fit other constraints, such as 
the closely-timed X-ray eclipse at conjunction (Pollock et al. 2005, fig. 1). 
If we drop the assumption that the acceleration of the plasma in the WCR was 
the same for all the dust features, the phase at which wind compression began 
to be sufficient for dust formation could have occurred earlier. There is no 
scope for a similar adjustment to the phase for the end of dust formation, as 
the proper motion `start date', assumed delay and P.A. on the sky are consistent.

As discussed in Section \ref{SStart}, the duration of dust nucleation was half 
the formation interval indicated by the proper motions. Unsurpisingly, model 
dust images (Fig.\,\ref{ModelComp}) using the nucleation interval for the 
duration of dust formation gave worse fits to the observed images than those 
using the duration from the proper motions. 

The simple model for dust formation in the WCR does not fully account for the 
dust observed around WR\,140 whatever the choice of input parameters, so the 
model assumptions need further examination. We have already found that 
a model including an asymmetric distribution of dust around the axis of the 
WCR improves the fit to the observed images slightly. 
There is a stronger piece of evidence supporting this non-uniform azimuthal 
density distribution around the WCR axis: the different proper motion of `C0' 
and `C1', considered (Section \ref{SPimages}) to be caused by the different 
inclination angles of their trajectories. The axis of the WCR was tilted away 
from us during the time when these features were formed, with the angle between 
the axis and our sightline ($\psi$, where $\cos(\psi) = -\sin(i)\sin(f+\omega)$, 
with $f$ being the true anomaly of the O5 star in its orbit) varying over 
$\psi=119\degr-104\degr$ for $\phi=0.015-0.025$. With $\theta = 40\degr$, this 
has dust flowing both towards us and away from us, and having a three-fold 
range of projected velocities on the sky. However, if the dust were distributed 
symmetrically around the WCR axis, because the emission is optically thin, 
the strongest dust emission would come from the limb-brightened edges of the 
cone, which would have the same projected distances from the star 
{\em whatever the orientation of the WCR cone}. In this case, we would be unlikely 
to observe the effects of different inclinations of the dust motions. This is 
confirmed by the modelling. Models of the dust emission for different phases 
assuming uniform azimuthal dust show that the concentrations to the south 
identified with `C0' and `C1' have the same proper motions. In the models with 
asymmetric dust, the proper motion of `C0' is found to be 70 per cent that of 
`C1', close to the observed value of 75 per cent (Section \ref{SMotion}).
The higher density in the following edge of the WCR accords with theoretical 
studies of WCRs in adiabatic systems; when the studies have been extended to 
systems like WR\,140, the dust modelling should be repeated with a more realistic 
azimuthal density distribution than the simple one used here. 

A similar asymmetry may apply to the dust made by the classical pinwheel WR\,104, 
but will be harder to observe. Because its dust formation is continuous, there 
are no concentrations to track by proper motion, and the dust plume is narrower 
than that of WR\,140 -- see, especially, fig.\,8 of Tuthill et al. (2008) 
for a stacked image -- so it is harder to observe any structure in the pinwheel.
It would be instructive to observe another pinwheel having a low orbital 
inclination, like that of WR\,104, but a more luminous companion to the WC9 star 
and hence larger values of $\eta$ and $\theta$ and a wider dust plume which 
might show a density gradient.

There is a second category of asymmetry in the dust formation by WR\,140: the 
four-fold difference in the amount of dust formed in the features before and 
after periastron, and the difference in time-scales. 
Either the wind-compression process depends on more than the local conditions 
-- radiation field and pre-shock wind density -- which depend on stellar 
separation only, or the difference in dust mass in the `arm' and `bar' must be 
a consequence of a significant difference in the efficiency of dust formation 
from the wind. 

We expect a delay between the compression of the dust-forming plasma and 
nucleation of the dust, to allow time for the plasma to flow to where the 
dust forms. If this is the nucleation radius ($\sim 125$~au), the delay would 
be about 0.03~P for a flow at about 2400~km~s$^{-1}$ (Section \ref{SMotion}). 
Such a delay is consistent with the beginning of nucleation at $\phi = 0.0$ 
(Section \ref{SPhot}) and the start phase, $\phi = 0.972$, adopted from the 
proper motions, so we could have dust nucleating as soon as it reached the 
nucleation radius. On the other hand, the cessation of nucleation between 
$\phi=0.025$ and $\phi=0.03$ at the latest places a tight limit on the delay, 
as the stop phase from the proper motions of `C0' and `C1', $\phi = 0.028$, 
and the P.A. of the `bar', which is consistent with the condensation of the 
dust-forming plasma at the same phase, leave no time for it to flow to the 
nucleation radius. 

If the basic model and the orbital parameters are correct, the time-scale 
for nucleation for the southern dust features suggests that dust is able to 
condense much closer to the stars than the radiative-equilibrium nucleation 
radius, perhaps because of screening by the compressed wind. Once some 
dust grains have condensed, they will themselves provide screening and 
facilitiate further condensation.
The models of line-of-sight dust clouds formed by WC9 stars (Veen et al. 1998) 
suggest that these clouds form closer to the stars than the persistent dust 
clouds located beyond the nucleation radii. Something similar may be 
happening in WR\,140, but the fact that this occurred only at the end of the 
dust-forming episode, and then produced most of the dust, suggests different 
processes governing the formation of the early and final dust features. 

Nevertheless, the basic features of dust formation by WR\,140 can be ascribed 
to the CWB paradigm, including the spread of dust by the rotating WCR, but 
the processes at the start and finish of the process -- which are not observed  
with the persistent dust formation by the WC9 stars - have still to be 
understood.

\section{Conclusions} 
\label{SConc} 

The images of WR\,140 observed at wavelengths between 2 and 18 $\umu$m using 
four different instruments on four different telescopes give a consistent 
view of the evolution of the dust formed around the time of the 2001 
periastron passage until $\phi = 0.56$. Dust is observed to be distributed 
most of the way round the star, more like a `splash' than a short arc of a 
pinwheel like those made by WC9 stars as one might have expected from the 
very small fraction of the period (0.055~P) taken up by dust formation. This 
results from the fast angular motion of the binary and projection of the WCR 
on the sky around periastron as a consequence of the highly eccentric orbit. 

The positions of three persistent dust concentrations were tracked, showing 
them to be moving away from the stars with constant proper motions. 
Extrapolation of the linear motions back to the stars indicates that one 
concentration, that to the E of the binary, was ejected about 145~d before 
the other two, both to the S of the binary and comprising the `bar'. 
The constancy of the proper motions in the presence of strong radiation 
pressure indicates that the dust grains must have been accelerated to 
constant drift velocities relative to the dust-forming plasma. 
Continued movement through the plasma would have been important for growth 
of the grains by implantation of C ions (cf. Zubko 1998) after nucleation 
ceased, so that the dust mass continued growing. 
The multi-band ($H$--[19.5]) light curves showed nucleation ceased around 
$\phi \simeq 0.025-0.030$ but dust mass increasing until $\phi \simeq 0.14$. 
After this phase, the dust mass began to fall, presumably due to destruction 
of grains by spallation. Photometry of the most persistent dust feature, the 
southern `bar', from the [3.99]-band images show its mass falling by a factor 
of about four over $\phi = 0.23-0.56$ but this is probably an overestimate 
owing to the likely loss of faint flux from reconstructed images.
Grain destruction at later phases must be less important, because we observed 
dust emission from the `bar' made in the previous, 1993, dust-making episode at 
12.5~$\umu$m and 4.68~$\umu$m at distances consistent with the proper motions 
of the concentrations made in the 2001 episode.
This shows that, to the S at least, the dust expands freely in a low-density 
void, presumably that blown by the stellar winds. The dust feature to the E, 
the `arm', is not as robust, and partly dissipates during our sequence of 
observations. Photometry of the images shows the dust mass in the `arm' to be 
significantly less than that in the southern `bar' including `C0' and `C1'.

Model images of the dust cloud were constructed with dust forming downwind in 
the WCR, and using the projected angular motion of the WCR calculated from the 
orbital elements. 
Comparison of the model images with the observations shows a reasonable match 
for the persistent dust features to the south of the stars, but differences to 
the E and NW. The effects of adjustments to model and orbital parameters were 
explored: only the phase at which the dust-forming plasma started to be produced 
and the orbital eccentricity significantly affected the fits to the E. 
The fit to the NW, where the models produced more dust than observed, could be 
improved by adopting an asymmetric distribution of material around the axis of 
the WCR, with highest density at the following edge of the WCR in the orbital 
plane. 
This asymmetry also accounts for the difference in proper motions of the two 
most persistent dust features and accords with theoretical work on colliding 
winds. 
It is the most promising development of the basic WCR dust-formation model.

The shorter duration of dust nucleation interval suggests that the delay between 
wind compression and dust formation was different for different features. 
Dust does not necessarily start nucleation when the wind reaches the nucleation 
radius, and may even condense closer to the stars. The nucleation of dust in WC 
winds is still not understood, but the dust formation by WR\,140 is consistent 
with this process occurring in a rotating WCR -- as inferred for the rotating 
`pinwheels' like WR\,104 and WR\,98a, for which stellar orbits have yet to be  
observed.

\section*{Acknowledgments}

The United Kingdom Infrared Telescope (UKIRT) is operated on Mauna Kea, 
Hawaii, by the Joint Astronomy Centre (JAC), on behalf of the Science and 
Technology Facilities Council (STFC) of the UK. 
The William Herschel Telescope (WHT) is operated on Observatorio del 
Roque de los Muchachos, La Palma, by the Isaac Newton Group (ING), on 
behalf of the STFC. 
The Telescopio Carlos S\'anchez (TCS) in the Observatorio del Teide is 
managed by the Instituto de Astrof\'{i}sica de Canarias.
We are very grateful to the Service Programme observers at UKIRT, the TCS 
and the ING for the Service observations of WR\,140 over the years. 
AFJM is grateful for financial assistance to NSERC (Canada) and FQRNT 
(Quebec). We are grateful to Peter Tuthill for a helpful referee's report.

\bsp
\label{lastpage}

\begin{thebibliography}{99}

\bibitem[AHS]{}  Allen D. A., Harvey P. M., Swings J. P., 1972, A\&A, 20, 333

\bibitem {} Andersen A.C., Loidl R., H\"ofner S., 1999, A\&A, 349, 243

\bibitem[HI]{} Arnal E. M., 2001, AJ, 121, 413

\bibitem[thins]{} Cant\'o J., Raga A. C., Wilkin F. P., 1996, ApJ, 469, 729

\bibitem[C2]{} Cherchneff I., Le Teuff Y. H., Williams P. M., Tielens A. G. G. M.,   
               2000, A\&A, 357, 572

\bibitem[ACAR]{} Colangeli L., Mennella V., Palumbo P., Rotundi A., Bussoletti E., 
                 1995, A\&AS, 113, 561

\bibitem[Ostar]{} Conti P. S., Alschuler W. R., 1971, ApJ, 170, 325

\bibitem[48a]{}  Danks A. C., Dennefeld M., Wamsteker W. M., Shaver P. A., 
                 1983, A\&A, 118, 301
 
\bibitem[VLBA]{} Dougherty S. M., Beasley A. J., Claussen M. J., Zauderer A., 
                 Bolingbroke N. J., 2005, ApJ 623, 447

\bibitem[vinf]{} Eenens P. R. J., Williams P. M., 1994, MNRAS, 269, 1082

\bibitem[]{} Eichler D., Usov V. 1993, ApJ, 402, 271

\bibitem[ZetPup]{} Eversberg T., Lepine, S, Moffat A. F. J., 1998, ApJ 494, 799

\bibitem[cold]{} Folini D., Walder R., 2002, in:  Moffat A. F. J., St-Louis N. (eds.), 
               ASP Conference Series 260, Interacting Winds from Massive Stars,  
	       (San Francisco: ASP), p.~605 

\bibitem[Braking]{} Gayley K. G., Owocki S. P., Cranmer, S. R., 1997, ApJ 475, 786 

\bibitem[GH74]{} Gehrz R. D., Hackwell J.A., 1974, ApJ 149, 619

\bibitem[Michelle]{} Glasse A. C., Atad-Ettedgui E. I., Harris J. W., 1997, 
                     Proc SPIE, 2871, 1197

\bibitem[HGG]{}  Hackwell J.A., Gehrz R.D., Grasdalen G.L., 1979, ApJ, 234, 133

\bibitem[3D104]{} Harries T. J., Monnier J. D. Symington N. H. Kurosawa R., 
                  2004, MNRAS, 350, 565

\bibitem[PHARO]{} Hayward T. L., Brandl B., Pirger B.,  Blacken C., 
                  Gull G. E., Schoenwald J., Houck J. R., 2001, PASP 113, 105

\bibitem[CW2]{} Hill G. M., Moffat A. F. J., St-Louis N., 2002, MNRAS, 335, 1069


\bibitem{} Lamontagne, R., Moffat A. F. J., Seggewiss W., 1984, ApJ, 277, 258

\bibitem[cori]{} Lemaster M. N., Stone J. M., Gardiner T. A., 2007, ApJ, 662, 582

\bibitem[CWB]{} L\"uhrs S., 1997, PASP, 109, 504
 
\bibitem[WR137]{} Marchenko S. V., Moffat A. F. J., Grosdidier Y., 1999, 
                  ApJ, 522, 433

\bibitem[WR140]{} Marchenko S. V., et al., 2003, ApJ, 596, 1295 (MM03)

\bibitem[IRAS]{} Marston A. P., 1996, AJ, 112, 2828

\bibitem[Clumps]{} Moffat A. F. J., Robert C., 1994, ApJ 421, 310

\bibitem{} Monnier J. D., 2003, Rep. Prog. Phys. 66, 789

\bibitem[WR98a]{} Monnier J. D., Tuthill P. G., Danchi W. C., 1999, ApJ, 525, L97 

\bibitem[nonT]{} Monnier J.D., Greenhill L. J., Tuthill P. G., Danchi W. C., 
                 2002a, ApJ 566, 399

\bibitem[MTD]{} Monnier J. D., Tuthill P. G., Danchi W. C., 2002b, ApJ, 567, 
                L137 (MTD)

\bibitem[IOTA3]{} Monnier J. D. et al., 2004, ApJ 602, L57

\bibitem[cont]{} Morris P. W., Brownsberger K. R., Conti P. S., Massey P., 
                 Vacca W. D., 1993, ApJ, 412, 327

\bibitem[INGRID]{} Packham C., et al., 2003, MNRAS, 345, 395 

\bibitem{} Panov K. P., Dinko D., 2002, IBVS 5177

\bibitem{} Panov K. P., Altmann M., Seggewiss W., 2000,  A\&A, 355, 677

\bibitem[PiD140]{} Pittard J. M., Dougherty S. M., 2006, MNRAS, 372, 801

\bibitem[Mods]{} Pittard J. M., Dougherty S. M., Coker R. F., O'Connor E., 
                      Bolingbroke N. J., 2006, A\&A, 446, 1001

\bibitem[Chandra]{} Pollock A. M. T., Corcoran M. F., Steven I. R., Williams P. M.,
                    2005, ApJ 629, 482
                    
\bibitem[vinf]{} Prinja R. K., Barlow M. J., Howarth I. D., 1990, ApJ, 361, 607

\bibitem[UIST]{} Ramsay Howat S. K. et al.,  2004, Proc SPIE, 5492, 1160

\bibitem[Ostars]{} Repolust T., Puls J., Herrero A., 2004, A\&A, 415, 349

\bibitem[SPMCID]{} Salas L., Cruz-Gonz\'alez I., Tapia M., 2006, RMAA, 42, 273
 
\bibitem[Grids]{} Smith L. J., Norris R. P. F., Crowther P. A. 2002, MNRAS 337, 1039

\bibitem[CWBth]{} Stevens I. R., Blondin J. M., Pollock A. M. T., 1992, ApJ, 386, 265

\bibitem[Sputt]{} Tielens A. G. G. M., McKee C. F., Seab C. G., Hollenbach D. J.,
                       1994, ApJ 431, 321

\bibitem[WR104]{} Tuthill P. G., Monnier J. D., Danchi, W. C. 1999, Nature, 398, 486

\bibitem[308]{} Tuthill P. G., Monnier J. D., Danchi W. C., Turner N. H. 2003, 
                 in: van der Hucht, K. A., Herrero A., Esteban C. (eds.), 
                 Proc. IAU Symposium No. 212, A Massive Star Odyssey, from Main 
                 Sequence to Supernova,  (San Francisco: ASP), p.~121
                 
\bibitem[WR104b]{} Tuthill P. G., Monnier J. D., Lawrance N., Danchi W. C., Owocki S. P., 
                   Gayley K. G., 2008, ApJ 675, 698

\bibitem[DustTh]{}  Usov V.V. 1991, MNRAS, 252, 49

\bibitem[He140]{} Varricatt W. P., Williams P. M., Ashok N. M. 2004, 
                  MNRAS, 351, 1307 (VWA)

\bibitem[WR121]{} Veen P. M., van Genderen A. M., van der Hucht K. A., Li A., 
                  Sterken C., Dominik C., 1998, A\&A, 329, 199

\bibitem[Atlas]{} Walborn N. R., Fitzpatrick E. L., 1990, PASP, 102, 379

\bibitem{} Walder R., Folini, D., 2002, in:  Moffat A. F. J., St-Louis N. (eds.), 
    	   ASP Conference Series 260, Interacting Winds from Massive Stars,  
	       (San Francisco: ASP), p.~595 

\bibitem{} Walder R., Folini, D., 2003, in: van der Hucht, K. A., Herrero A., Esteban C. 
                 (eds.), Proc. IAU Symposium No. 212, A Massive Star Odyssey, from Main 
                 Sequence to Supernova,  (San Francisco: ASP), p.~139

\bibitem[VLA]{} White, R. L., Becker, R. H., 1995, ApJ 289, 698

\bibitem[IAU193]{} Williams P. M.  1999, in:  van der Hucht K. A., Koenigsberger  G.,
               Eenens P. R. J. (eds.), Proc. IAU Symp. No. 193, Wolf-Rayet Phenomena 
               in Massive Stars and Starburst Galaxies,  (San Francisco: ASP), p.~267

\bibitem{} Williams P. M.  2002, in:  Moffat A. F. J., St-Louis N. (eds.), 
	   ASP Conference Series 260, Interacting Winds from Massive Stars,  
	   (San Francisco: ASP), p.~311

\bibitem[Paper 0]{} Williams P. M., Beattie D. H., Lee T. J., Stewart J. M., 
                   Antonopoulou E., 1978, MNRAS, 185, 467

\bibitem[WHT]{} Williams P. M., van der Hucht K. A., Th\'e P. S., 1987, A\&A 182, 91.

\bibitem[Paper1]{} Williams P. M., van der Hucht K. A., Pollock A. M. T., 
                    Florkowski D. R., van der Woerd H., Wamstecker W. M.,  1990, 
                    MNRAS, 243, 662 (Paper 1)

\bibitem[SWS]{} Williams P. M., van der Hucht K. A., Morris P. W., 1998, ApSS, 255, 169

\bibitem{} Zhekov S. A., Skinner S. L., 2000, ApJ, 538, 808

\bibitem{} Zubko V. G., 1998, MNRAS, 295, 109

\bibitem{} Zubko V. G., Mennella V., Colangeli L., Bussoletti E., 1996,
            MNRAS, 282, 1321

\end{thebibliography}
\end{document}